\documentclass{aa}

\usepackage{graphicx}
\usepackage{txfonts}
\begin{document}

    \title{$BVRIJHK$ photometry and proper motion analysis of \object{NGC~6253} and the surrounding field.
     \thanks{Based on observation made at the European Southern Observatory, La
      Silla, Chile and at the Anglo-Australian Observatory,
      Siding Spring, Australia. 
      The catalog presented in this paper is only available 
      in electronic form at the CDS via anonymous ftp to 
      cdsarc.u-strasbg.fr (130.79.128.5) or via http://cdsweb.u-strasbg.fr/cgi-bin/qcat?J/A+A/ .}}

  \author{M. Montalto
          \inst{1,2},
          G. Piotto
          \inst{3},
          S. Desidera
          \inst{4},
          I. Platais
          \inst{5},
          G. Carraro,
          \inst{6}
          Y. Momany
          \inst{4},
          F. De Marchi
          \inst{3,8},\\
          A. Recio-Blanco
          \inst{7}.
          }

   \authorrunning{M. Montalto et al.}

   \offprints{M. Montalto,\\
              \email{montalto@usm.uni-muenchen.de} }

   \institute{Universitaets-Sternwarte  der
              Ludwig-Maximilians-Universitaet, Scheinerstr.1, 81679 Muenchen, Germany.
              \and
              Max-Planck-Institute for Extraterrestrial Physics, Giessenbachstr.,
              Garching b Muenchen, 85741, Germany.
             \and
              Dipartimento di Astronomia, Universit\`a di Padova,
              Vicolo dell'Osservatorio 2, I-35122, Padova, Italy
              \and
              INAF -- Osservatorio Astronomico di Padova,
              Vicolo dell' Osservatorio 5, I-35122, Padova, Italy
              \and
              Department of Physics and Astronomy,The Johns Hopkins University,
              3400 North Charles Street, Baltimore, MD 21218, USA
              \and
              ESO, Casilla 19001, Santiago 19, Chile
              \and
              Laboratoire Cassiop\'ee, Universit\'e de Nice Sophia-Antipolis,
              CNRS UMR 6202, Observatoire de la C\^ote d'Azur
              \and
              Dipartimento di Fisica, Universit\'a di Trento, 
              Via Sommarive 14, 38123 Povo(TN), Italy
             }

   \date{}

\abstract
{We      present    a       photometric         and        astrometric  catalog
of                $187963$                    stars              located in the
field    around    the old super-metal-rich  Galactic   open  cluster \object{NGC~6253}.
The           total         field-of-view       covered   by the
catalog          is           $34\arcmin\times33\arcmin$.   In this field,
we provide CCD $BVRI$ photometry.
For a  smaller region  close  to the cluster's center,  we also provide
 near-infrared  $JHK$ photometry.
}
{We analyze the properties of   \object{NGC~6253} by using our new photometric
data and astrometric membership.
}
{In June 2004, we targeted   the cluster during a $10$ day multi-site campaign,
which involved the MPG/ESO $2.2$m telescope with its wide-field imager and
the Anglo-Australian $3.9$m telescope, equipped with the IRIS2 near-infrared
imager. Archival CCD images of \object{NGC~6253} were used to derive relative proper
motions and to calculate the cluster membership probabilities.
}
{We have refined the cluster's fundamental parameters, deriving 
$(V_0-M_v)=11.15$, $ E(B - V)\,=\,0.15$,  $E(V - I)\,=\,0.25$,
$E(V - J)\,=\,0.50$, and   $E(V - H)\,=\,0.55$.
The     color     excess    ratios  obtained using  both
the optical and near    infrared    colors   indicate   a  normal reddening
law in the direction of \object{NGC~6253}.
The age of \object{NGC~6253} at 3.5~Gyr, determined from our best-fitting isochrone
appears to be slightly older than the previous estimates. Finally, we
estimated the binary fraction among the cluster members to be  $\sim\,20\%-30\%$
and identified $11$ blue straggler candidates.
}
{}


   \maketitle

%

\section{Introduction}
\label{s:introduction}

Studies of old Galactic open clusters at a range of different ages,
metallicities, and spatial location in the disk are fundamental for
understanding the formation and chemical/dynamical evolution
of the Galactic disk since its early assembly (Janes et al. 1994,
Friel et al. 2002, Bragaglia \& Tosi 2006).   As groups of stars sharing
common metal abundance, age, and distance, the star clusters are ideal
tracers because some of their fundamental parameters can be derived
on statistical basis.

Over the years most attention has been dedicated to the solar
neighborhood, where the properties of open clusters can be considered
relatively complete for up to 1~kpc    from   the Sun (Piskunov et al. 2006).
The open clusters in the direction of  Galactic center and anticenter are,
however, poorly studied.      Recent attempts by Yong et al. (2005)
and Carraro et al. (2007) seem to indicate that the  age-metallicity
relationship and the radial abundance gradient toward the
Galactic anticenter do not deviate substantially
from that in the solar neighborhood.   The Galactic region inside the solar
ring is more difficult to study because the interstellar absorption is
stronger and the higher-density stellar environments shorten the cluster
survival time. Nevertheless, a few open clusters of unexpectedly old age
have been detected toward the Galactic center (Carraro et al. 2005a, 2005b),
which allows to extend the baseline for the Galactic radial abundance
gradient studies by up to 2-3 kpc in the bulge direction, once the metal
abundances are measured.

The two old and extremely metal-rich open clusters, \object{NGC~6791}
and \object{NGC~6253}, also located in the inner Galaxy, have been studied
relatively extensively. Such metal rich clusters are more likely to
harbor extrasolar planets, following the well-established correlation
between the frequency of giant planets and the metallicity of a host
star (e.g., Valenti \& Fischer 2008). 
Searching for extrasolar planets in metal-rich open clusters should thus
increase the probability of planetary detection, allowing us to study
the planetary formation and evolution processes among the stars that share
well-defined common properties. Although photometric monitoring using
the transit techniques has yet to produced bona fide extrasolar planets
in open clusters, these observations have provided constraints
on the expected frequency of extrasolar planets (e.g., Hartman, et al. 2008, 
Aigrain et  al. 2007, Weldrake et al. 2005, Street et al. 2003, Burke et al. 2006,
Mochejska et al. 2006).

In 2002 we started a project to search for transiting  planets in
metal-rich open clusters. First, we focused our attention on \object{NGC~6791},
the most populous, metal-rich open cluster known in our Galaxy (Montalto
et al. 2007, De Marchi et al. 2007). This ten consecutive night monitoring
of \object{NGC~6791} produced the null-detection of transits (Montalto
et al. 2007).  Then, we chose the other metal-rich open cluster,
\object{NGC~6253} ($\alpha_{2000}=16^{h}\,59^{m}\,05^{s}$, $\delta_{2000}=-52^{\circ}\,42\arcmin\,30\arcsec$,
$l=335\fdg5,b=-6\fdg3$), for planetary
transit searches.
We obtained photometric data in seven photometric bandpasses, ranging from
optical $B$ to near-infrared $K$ filters, during a $10$ days multi-site
campaign at the MPG/ESO $2.2$m telescope (La Silla, Chile) and at the
Anglo-Australian $3.9$m telescope (Siding Spring) in June 2004.
This represents the largest photometric dataset obtained for this
cluster as shown in Table\ref{tab:results}.
We compare our results with those
obtained       from    past    photometric and spectroscopic studies of this cluster.

To date, \object{NGC~6253} has already been studied photometrically by five
groups: Bragaglia et al.~(\cite{bragaglia97}), Piatti et al.~(\cite{piatti98}),
Sagar, Munari, \& de Boer~(\cite{sagar01}), Twarog, Anthony-Twarog
\& De Lee~(\cite{twarog03}), Anthony-Twarog, Twarog, \& Mayer (2007), 
hereafter referred to as BR, PI, SA, TW, and AT, respectively.
Another three  studies by Carretta et al. (2000), Carretta, Bragaglia
\& Gratton (2007), and Sestito et al. (2007), (hereafter referred to as
CA1, CA2, and SE) have primarily established the super-high metallicity
of \object{NGC~6253} from high-resolution spectroscopy. A summary of
the derived fundamental parameters from these studies is given in Table~1.
The basic parameters of \object{NGC~6253} show a rather wide spread from
study to study, suggesting the following mean parameters:
reddening $E(B-V)$=0.2, true distance modulus $(V-M_v)_0$=10.9,
metallicity [Fe/H]=$+0.3$ dex, age $\sim$3~Gyr. In addition, from
CA1, CA2, and SE papers we estimate the mean heliocentric radial
velocity of \object{NGC~6253} to be at $-28.0$ km~s$^{-1}$, with the large
4 km~s$^{-1}$ rms scatter.

Being projected against the Galactic Bulge area, this cluster is highly
contaminated by field stars. This limits substantially the accuracy of
derived cluster parameters, therefore one of our tasks was to obtain
proper motions and calculate reliable cluster membership probabilities.
The kinematic data then allow to isolate the cluster members in
color-magnitude diagrams and improve the analysis of cluster parameters.
Our primary motivation of this study was to assist the search for
transiting planets and its subsequent analysis.
We plan to report on the results of our transit search campaign and
variability  studies  in the forthcoming papers.


This   paper is organized as follows: in Sec.~\ref{s:observations} we describe
the observational material; in Sec.~\ref{s:reduction} we
give a detailed description of reductions and calibrations. Comparison of
our new photometry with the previous photometric studies is discussed
in Sec.~\ref{s:comparisons}.  The astrometric calibration is presented
in  Sec.~\ref{s:astrometric}.  Proper motions and cluster membership
analysis are discussed in Sec.~\ref{s:propermotion}, while the resulting
color-magnitude diagrams are presented in Sec.~\ref{s:cmd}.

\begin{table*}
\caption{
Fundamental parameters of \object{NGC~6253} from prior photometric and
spectroscopic studies.
\label{tab:results}
}

\begin{center}
\begin{tabular}{c c c c c c}
\hline
 Source   & $E(B-V)$ & $(m-M)_{0}$ & Age (Gyr) & [Fe/H] &  Aperture (m)\\
\hline
 BR  & $0.23$--$0.32$ & $10.9\,\pm\,0.1$ & $\sim\,3$ & $\sim+0.3\,$ & $1.5$\\
 PI  & $0.20\,\pm\,0.05$ & $10.90\,\pm\,0.40$ & $5.0\,\pm\,1.5$ & $+0.2$ & $0.61$\\
 SA  & $0.2$ & $11.30\,\pm\,0.15$ & $2.5\,\pm\,0.6$ & $+0.4$  & $1$\\
 TW  & $0.260\,\pm\,0.003$ & $10.8$--$11.3$ & $2.5$--$3.5$ & $\sim+0.4-0.7$\footnote{uno} & $0.9$\\
 AT  & $0.16\,\pm\,0.02$ &  &  & $+0.58\,\pm\,0.01$ & $0.9$\\
 CA1 & & & & $+0.36\,\pm\,0.20$ & \\
 CA2 & & & & $+0.46$ & \\
 SE  & & & & $+0.36\,\pm\,0.07$ & \\
\hline
\multicolumn{6}{c}{{\footnotesize$^{1}$ The low estimate is based on
isochrone fit, the high estimate is from the analysis of photometric indices.}}\\
\end{tabular}
\end{center}
\end{table*}

\section{Observations}
\label{s:observations}

\subsection{La Silla} A total of ten consecutive nights,
from June $14$ to June 23, 2005 at the MPG/ESO 2.2m telescope,
were dedicated completely to the observations of \object{NGC 6253}. 
We  used  the wide field imager (WFI)  which  consists of a mosaic of eight
CCDs $\rm 2K\,\times\,4K$, providing a field-of-view of  $\rm\,34'\,x\,33'$.
The pixel scale of WFI is   $0.238$   arcsec pixel$^{-1}$. The cluster was
centered on CCD No.  51, according to WFI nomenclature.
In total, we used $\sim$45.3$\,$ observational hours for \object{NGC~6253},
mainly in the $R$ filter. We also acquired the CCD images of \object{NGC~6253}
in $BVI$ filters along with the photometric standards (Landolt 1992).
The total number of CCD images with the cluster 918.

During the 10-day observing run the seeing (as measured directly from our images)
ranged between $0.5\arcsec$ to $2\arcsec$
with the mean seeing at $1\arcsec$ (Table~\ref{tab:LaSillaObs}).
On average, there was a trend of the worsening seeing with time 
over the entire observing run. To maintain the same S/N ratio, the
exposure time was adjusted accordingly, if the seeing deteriorated.
Bearing in mind the goal of obtaining high precision photometry
for searches of extrasolar planetary transits and stellar variability,
we always kept the stars at the fixed pixel positions, in order to minimize
the flat-fielding errors. More details about our La Silla observing runs
are given in Table~\ref{tab:LaSillaObs}, Table~\ref{tab:LaSillaObsbvi}.
In Table~\ref{tab:LaSillaObs}, we present the observations in $R$ band.
Each night is indicated by a Roman numeral in the first column. The following 
columns indicate: the starting and ending UT times of observation,
the number of observing hours calculated
as a sum of all exposure times, the mean seeing measured from the images
and its standard deviation
in parenthesis, the range of exposure times and the 
total number of collected images (N.imm.).
In Table~\ref{tab:LaSillaObsbvi}, we present the observations acquired
in $B$, $V$ and $I$ bands. The night of observation is indicated in the 
first column and in the following we show the number of images obtained 
and their exposure times for each filter.

\subsection{Siding Spring}
A total of 10 observing nights for this program were also allocated
at the Anglo-Australian 3.9m Telescope (AAT), essentially concurrent with La Silla
observations.  Originally, the observations were scheduled with the wide
field imager but because of a detector failure we decided to use
the IRIS2 infrared detector instead. This is a 
$1K\,\times\,1K$ detector with a $18.5\,\times\,18.5$ micron pixels,
 a field-of-view of $7.7'\,\times\,7.7'$ and a $0.45$ arcsec pixel$^{-1}$
scale. Due to the inclement weather and scheduling problems, only three
nights were used to obtain the cluster images in $J$ filter
(Table~\ref{tab:SidingSpringObs}). A total of 60 images in $H$ and $K$ filters
were obtained during the last night (number IX). The seeing varied from
1.7 to 1.1 arcsec in the $J$ band, as measured directly from our images. 

The image acquisition strategy was different for Siding Spring.
Because of variable background for the IRIS2 detector, following each
10 cluster images it was necessary
to acquire a field image, in order to
correct for this effect. Also, the exposure time for \object{NGC~6253}
was always fixed at $60$~s (apart for a few short exposure images).
In addition, frequent shifts between the field and the cluster
deteriorated the precision of the telescope's pointing and guiding,
which affected the quality of images. 
Additional details about our near-infrared $JHK$ observations are
provided in Tables~\ref{tab:SidingSpringObs},\ref{tab:SidingSpringObshk}.

\begin{table*}
\caption{
Journal of observations for \object{NGC~6253} taken in June $14$-$23$, $2004$ at
La Silla in $R$ filter. 
\label{tab:LaSillaObs}
}
\begin{center}
\begin{tabular}{c c c c c c c}
\hline
 Night & UT(start) & UT(end) & Hours & Seeing (arcsec) & Exp (s) & N.imm. \\
\hline
    I    &   1:40  &   9:47   & 4.16  &  0.9(0.2) &  70-600 &  86 \\
   II    &  23:46  &   9:50   & 5.28  &  0.7(0.2) &   3-400 & 143 \\
  III    &  23:06  &   9:31   & 5.79  &  0.8(0.1) &  70-900 & 126 \\
   IV    &   4:26  &   8:54   & 3.92  &  1.7(0.2) & 300-900 &  24 \\
    V    &  23:29  &   9:26   & 4.06  &  0.8(0.1) &   3-185 & 142 \\
   VI    &  23:20  &   9:35   & 5.83  &  1.0(0.2) &  90-400 & 154 \\
  VII    &  22:58  &   5:08   & 4.01  &  1.7(0.2) & 100-900 &  40 \\
 VIII    &   7:58  &   9:05   & 1.75  &  1.9(0.2) & 100-900 &   6 \\
   IX    &  22:32  &   6:29   & 4.67  &  1.4(0.2) & 100-900 &  61 \\
    X    &  23:03  &   9:35   & 5.81  &  1.2(0.2) & 100-650 & 124 \\
\hline
\end{tabular}
\end{center}
\end{table*}

\begin{table*}
\caption{
Journal of observations for La Silla in $B$, $V$, and $I$ filters.
\label{tab:LaSillaObsbvi}
}
\begin{center}
\begin{tabular}{c c c c c c}
\hline
 Nights & $B$ & $V$ & $I$ \\
\hline
 I, V & $1$x$10s$, $2$x$185s$ & $2$x$6s$, $2$x$160s$, $2$x$120s$ & $1$x$2s$, $2$x$100s$ &  & \\
\hline
\end{tabular}
\end{center}
\end{table*}

\begin{table*}
\caption{
Journal of observations for \object{NGC~6253} taken on $15$, $17$ and $22$ of June,
$2004$ at Siding Spring in $J$ band. 
\label{tab:SidingSpringObs}
}

\begin{center}
\begin{tabular}{c c c c c c c}
\hline
 Night & UT(start) & UT(end) & Hours & Seeing (arcsec) & Exp (s) & N.imm. \\
\hline
   II    &  12:30  &   18:30   & 3.52  &  1.7(0.3) & 5-60 & 218 \\
   IV    &   7:52  &   16:25   & 5.80  &  1.5(0.4) & 60 & 348 \\
   IX    &   8:48  &   18:39   & 6.72  &  1.1(0.3) & 10-60 & 348 \\
\hline
\end{tabular}
\end{center}
\end{table*}

\begin{table*}
\caption{
Journal of observations for Siding Spring in $H$ and $K$ filters.
\label{tab:SidingSpringObshk}
}

\begin{center}
\begin{tabular}{c c c}
\hline
  Night & $H$ & $K$ \\
\hline
  IX & $30$x$10s$ & $30$x$10s$ \\
\hline
\end{tabular}
\end{center}
\end{table*}

\section{Reduction of the data}
\label{s:reduction}

\subsection{Pre-reduction}
The pre-reduction steps were performed using the common IRAF
\footnote{IRAF is distributed by the National Optical Astronomy Observatory, which
is operated by the Association of Universities for Research in Astronomy, Inc.,
under cooperative agreement with the NSF.} software system.
 For the Siding Spring images, we used a different approach to obtain the
flat-field corrections. 
 
For each night, we built a master sky frame from the dithered field images
taken around the cluster as described in Section~\ref{s:observations}. 
For each image we estimated the mean sky 
level in manually-selected window
free of stars, and scaled the images to the same mean count level. Then we
used the IRAF task $combine$ to create a master frame, using a 
3-$\sigma$ clipping algorithm. We also imposed an upper cut-off limit
to the pixel values to be considered in calculating the mean at
$\sim5-\sigma$ above the mean sky value, in order to effectively 
remove the sources from the sky frame. Then the resulting image
was normalized and all other images taken during the same night were
divided by it. We obtained 43, 41 and 35 field images in the $J$ filter
for the three observing nights shown in Table~\ref{tab:SidingSpringObs}.
During the last night we also acquired 5 field images in $H$ and
$K$ filters.

\subsection{Profile fitting photometry}

 We reduced the two datasets from La Silla and Siding Spring separately,
although we followed the same procedure for both of them. First,
we generated a PSF for each image in each dataset with
DAOPHOT II (Stetson 1987).  The analysis of La Silla dataset
 was performed separately for each chip of the MPG/ESO WFI imager.
 We chose the second-order, spatially variable PSF
 for La Silla images, while after some tests for the Siding Spring ones
 we decided to apply a constant PSF. The PSF was calculated via
 an iterative process, which progressively eliminates the stars
 which highest residuals. We verified
 that after 3-4 iterations the PSF does not change significantly, hence
 we chose to apply 4 iterations for each image.

Once the PSFs were obtained, we applied ALLSTAR to refine the stellar
 positions and magnitudes and then DAOMATCH and DAOMASTER to calculate the
 coordinate transformations between the frames. To do this, we chose
the best image in each dataset as a reference.
By stacking 50 best-seeing images with MONTAGE2, we calculated the
reference master frame from which we derived a master list of stars.
We note that the center of \object{NGC~6253} is located on CCD chip No. 51.
For the reductions of this particular CCD, we considered all images acquired
in $R$ filter and also those acquired in the other filters. Then,
we applied ALLFRAME (Stetson 1994) to the whole dataset. In the case of remaining WFI
CCD chips we considered only the best-seing images in $R$ filter. 
For the Siding Spring images, ALLFRAME was applied to the whole dataset.

\begin{figure}
\center
\includegraphics[width=9cm]{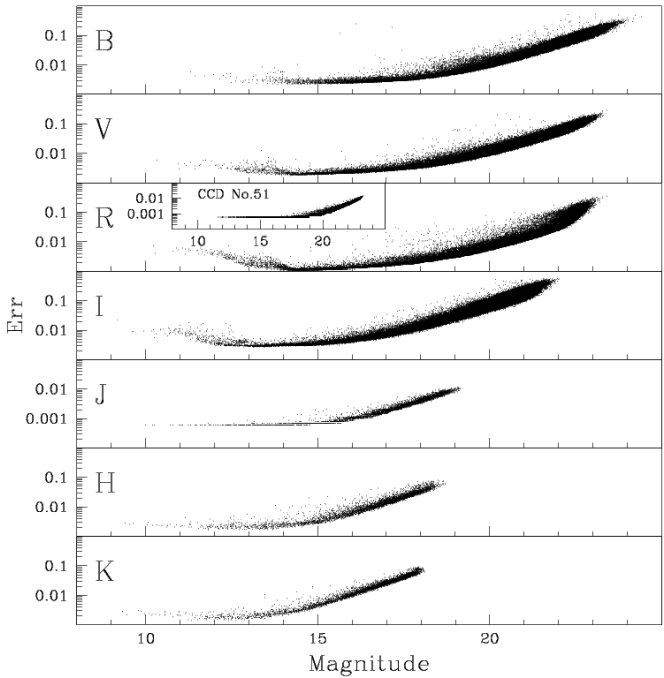}
    \caption{Standard error of the mean magnitudes as  measured in
             different photometric bandpasses.
        }
\label{fig:internal}
\end{figure}

\begin{figure}
\center
\includegraphics[width=9cm]{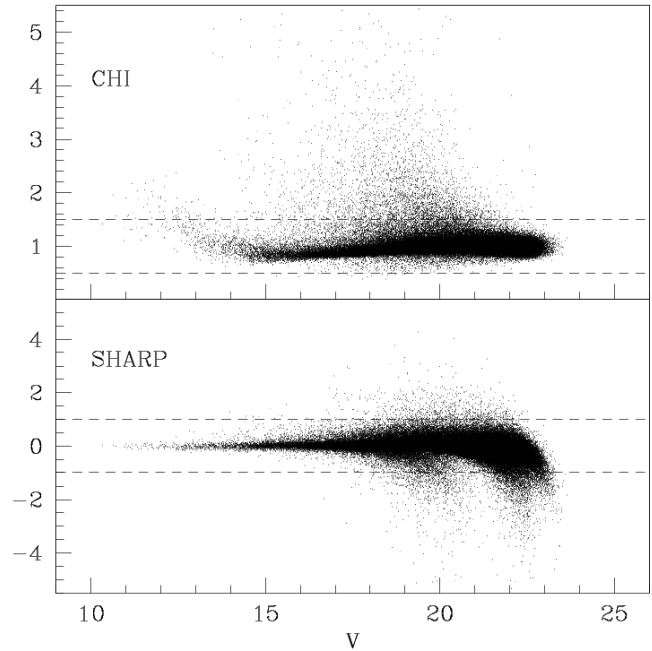}
    \caption{Fit parameters $\chi$ and $sharp$ and their range for the selection of
catalog objects.}
\label{fig:chsh}
\end{figure}

 We averaged the magnitudes in the same filter, if a star was present at 
least in half of the images. Figure~\ref{fig:internal} shows the
standard error of the mean magnitudes in our photometry, measured in
different filters. These errors are smaller in $R$ and $J$ filters
because the majority of images were taken through these filters. 
A preliminary selection of stars in the catalog was performed
using the goodness parameters of the fit, $\chi$ and $sharp$, provided
by ALLFRAME task and shown in Fig.~\ref{fig:chsh}.
We considered only those stars with the absolute value of sharpness
less than $1$, and the $\chi$ values comprised between $0.5$ and $1.5$.
The vast majority of eliminated sources are spurious detections,
faint objects close to bright images, or bright stars for
which reliable photometry is not possible due to the saturation.
The final catalog contains 187963 stars down to $V\sim$23.5 mag.

\subsection{Calibration}
\label{s:calibration}

\subsubsection{La Silla}
\begin{figure}
\center
\includegraphics[width=8cm]{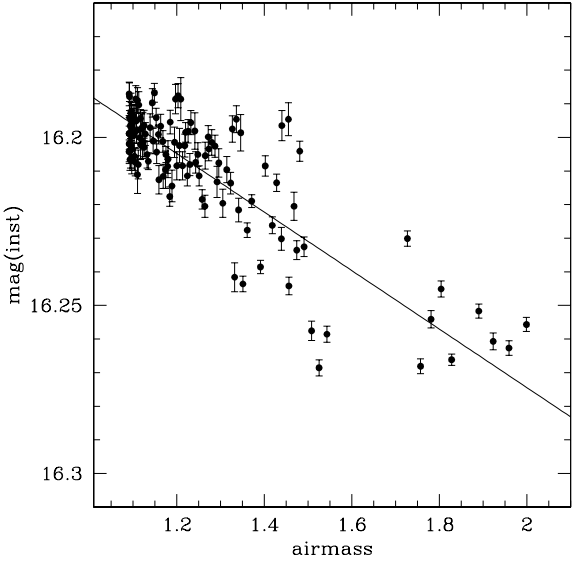}
    \caption{A plot of exposure-time-corrected instrumental $R$ magnitudes
of a bright  isolated star versus the airmass, all obtained during
the night of photometric calibrations at La Silla. The
             continuous line represents the best-fit weighted least square
             linear  model to the data, implying an extinction coefficient
             in  $R$ filter to be $0.09\,\pm\,0.02$ mag per unit airmass.
             The $rms$ of the fit is $\sim$0.02~mag.
        }
\label{fig:bouguer}
\end{figure}

The  calibration  of La Silla dataset was performed separately for each
CCD.
The $BVI$  photometric standard field Mark~A  was observed in
different filters on night V (Table~\ref{tab:LaSillaObsbvi}).
In  order  to check the photometric
stability of the night, we used our photometric sequence obtained in $R$
filter. A bright and isolated star close to the cluster center was
selected. Its measured instrumental magnitude was scaled to a $1\,$s
of exposure time and  plotted against the airmass (Fig.~\ref{fig:bouguer}).
By definition, a weighted least squares linear fit to the observed $R$
magnitudes as a function of airmass is the mean extinction coefficient
(slope) in this filter. For night V, the derived extinction
cofficient is $0.09\,\pm\,0.02$ mag per unit of airmass, consistent
with the   value reported on the $ESO$ WFI instrument webpage
for this filter
\footnote{http://www.ls.eso.org/lasilla/sciops/2p2/E2p2M/WFI/zeropoints/}
($0.07\,\pm\,0.01$).
We checked for the consistency of extinction across the entire field-of-view
by measuring the other isolated bright stars.
The lack of measurements around the airmass of $1.65$ (Fig.\ref{fig:bouguer})
is due to the fact that we then switched to the observations of the
standard field.

A total of $71$ standard stars were used to derive the calibration
coefficients.  The exposures  were   taken in all filters shown in
Tables~\ref{tab:LaSillaObs},\ref{tab:LaSillaObsbvi}.  The $BVRI$
photometry  of  the  standard  stars was taken from the database of Stetson
Photometric Standard fields~\footnote{http://cadcwww.dao.nrc.ca/standards/}.
The range of standard star colors spans  $-0.3<B-V<1.5$,
$0.6<V-I<2.3$     and     $0.3<V-R<1$,  accordingly, and thus is fully
adequate to calibrate our photometry.

For   the  standard field, the instrumental magnitudes were derived using
aperture photometry.  These magnitudes were then adjusted to $1$~s
of exposure time and corrected for atmospheric extinction.
The aperture corrections were applied out up to $6$ arcsec from
the stellar centroids. Here, we report the calibration equations
obtained for CCD No.~51:

\begin{equation}
 B-b=0.207(\pm0.007)(B-V) + 24.724(\pm0.005)
\end{equation}

\begin{equation}
 V-v=-0.089(\pm0.004)(B-V) +     24.065     (\pm  0.004)
\end{equation}

\begin{equation}
 V-v=-0.098   (\pm    0.006)(V-R) +     24.043    (\pm   0.003)
\end{equation}

\begin{equation}
 R-r=-0.083   (\pm    0.006)(V-R)  +    24.431    (\pm   0.003)
\end{equation}

\begin{equation}
 R-r=-0.075   (\pm    0.006)(R-I)  +    24.427    (\pm   0.003)
\end{equation}

\begin{equation}
 I-i=0.265   (\pm    0.007)(R-I)  +    22.924    (\pm   0.004)
\end{equation}

\begin{equation}
 B-b=0.119   (\pm    0.004)(B-R)  +    24.738    (\pm   0.005)
\end{equation}

\begin{equation}
 R-r=-0.023   (\pm    0.003)(B-R)  +    24.420    (\pm   0.004)
\end{equation}

\begin{equation}
 B-b=0.088    (\pm   0.003)(B-I)  +    24.733    (\pm   0.005)
\end{equation}

\begin{equation}
 I-i=0.076    (\pm   0.003)(B-I)  +    22.916    (\pm   0.005)
\end{equation}

\begin{equation}
 V-v=-0.063   (\pm    0.003)(V-I)  +    24.057    (\pm   0.004)
\end{equation}

\begin{equation}
 I-i=0.133   (\pm    0.005)(V-I)  +    22.920     (\pm  0.005).
\end{equation}

\subsubsection{Siding Spring}


The calibrations of the Siding Spring near-infrared observations were
obtained using the 2MASS photometry (Skrutskie et al., 2006).
We calculated the weighted mean differences between our own photometry and
the $2MASS$, using as weights the quadratic sum of our own photometric errors and those of the 
$2MASS$, and considering only the brightest sources ($J<14$). In total we used
270, 244 and 246 stars for $J$, $H$ and $K$ band respectively. The result is
reported in the following equations:

\begin{equation}
J_{AAT}-J_{2MASS}=(0.06\pm0.08)
\end{equation}

\begin{equation}
H_{AAT}-H_{2MASS}=(-0.06\pm0.07)
\end{equation}

\begin{equation}
K_{AAT}-K_{2MASS}=(0.40\pm0.16).
\end{equation}

Our photometry is equivalent to the $2MASS$ in the $J$ and $H$ band, but 
appears slightly poorer in the $K$ band, as indicated also by the larger offset and 
the $\sigma$ reported above. This is probably due to the small number of
 scientific images acquired
and too few background images used for the sky removal. For this reason the $K$ band 
is not further used in this work.
Moreover, in our final catalog we report both our own infrared photometry and the $2MASS$
photometry, for the $1017$ stars in the cluster region in common to our catalog which have at least 
one measurement in $J$, $H$ or $K$ band.

\section{Comparison with previous photometry}
\label{s:comparisons}
In previous photometric studies of \object{NGC~6253}, the broadband CCD
data were obtained in $UBVRI$ filters (studies by BR and SA) and in $BVI$
for PI, while TW presented CCD photometry on the
intermediate-band $ubvyCa$H$\beta$ system. Systematic differences in
the broadband data have been already pointed out by PI and TW.
In order to compare our data with those in the other surveys,
we calculated the differences ``our minus other photometry".
These residuals were fit with a linear function in the form $a(col)+b$,
where $col$ is the chosen color index.
The coefficients $a$ and $b$ along with their errors ($\sigma_a$,
$\sigma_b$),the dispersion of the residuals ($rms$), the number of
stars used in the fit ($N$), and the number of rejected stars
($N_{\rm rej}$) are reported in Table~\ref{tab:confphot}.
The differences with the other broadband data from BR, SA and PI,
imply zero point offsets in $V$ filter ranging between 0.01-0.1 mag and
color terms  ranging between $0.03-0.07$ mag. On average, our photometry
is closer to that by SA, although it varies and depends on the
bandpass considered. In the following discussion we decided to use
only our own photometry, because of the data homogeneity
and the high photometric quality of a calibration night, as indicated by
Fig.~\ref{fig:bouguer}.


\begin{table*}
\caption{Comparison with previous broadband photometric studies.
\label{tab:confphot}
}

\begin{center}
\begin{tabular}{c c c c c c c c c}
\hline
 Survey & Color &  $a$ & $\sigma_{a}$ & $b$ & $\sigma_{b}$ & $rms$ & $N$ & $N_{rej}$ \\
\hline
  BR($B$) & $B-V$ & 0.05   & 0.01  &-0.13 &  0.01 & 0.02 & 296 & 4 \\
  BR($V$) & $B-V$ & 0.07   & 0.01  &-0.11 &  0.01 & 0.02 & 295 & 5 \\
  BR($R$) & $V-R$ & 0.003  & 0.010 & 0.02 &  0.01 & 0.02 & 296 & 4 \\
  BR($I$) & $V-I$ & -0.02   & 0.01  & 0.06 &  0.01 & 0.02 & 295 & 5 \\
\hline
  SA($B$) & $B-V$ &-0.08  & 0.03  & 0.07 &  0.02 & 0.05 & 95 & 1 \\
  SA($V$) & $B-V$ & 0.03  & 0.02  & 0.02 &  0.02 & 0.03 & 93 & 3 \\
  SA($R$) & $V-R$ & -0.04  & 0.04  & 0.15 &  0.02 & 0.04 & 94 & 2 \\
  SA($I$) & $V-I$ &  0.007 & 0.020 & 0.05 &  0.02 & 0.04 & 93 & 3 \\
\hline
  PI($B$) & $B-V$ & -0.17  & 0.04 &  0.06  & 0.03  & 0.07  & 134 & 1 \\
  PI($V$) & $B-V$ &  0.03  & 0.01 & -0.05  & 0.01  & 0.03  & 131 & 4 \\
  PI($I$) & $V-I$ & 0.01  & 0.02 & -0.003 & 0.020 & 0.04  & 131 & 4 \\
\hline
\end{tabular}
\end{center}
\end{table*}

\section{Astrometric calibration}
\label{s:astrometric}
The astrometric calibration of pixel coordinates was     performed by means of the IRAF
subroutines,       using     the     UCAC2 (Zacharias et al. 2004)
 as a reference catalog.
A trial solution for each frame was obtained from a manually selected
sample  of UCAC2 stars in the field.
The so-called equatorial solution  was      found by      means of the IRAF
tasks   $ccdmap$ of $ccdtran$. An iterative procedure was then applied
to        refine        the        coordinates   by   including all
available reference stars from UCAC2. This
procedure     was   applied separately to each one of the CCDs in the
MPG/ESO WFI imager. The Siding Spring positional data were calculated
using the equatorial coordinates from the La Silla final catalog of
positions in \object{NGC 6253}. The accuracy of astrometric calibration
in our catalog is $\leq0.1\arcsec$ in both coordinates.

\section{Proper motions}
\label{s:propermotion}

\object{NGC~6253} is projected towards a very rich stellar field at a low
galactic latitude, making the identification of cluster members from
photometry alone highly uncertain. Proper motions is a powerful and
reliable tool to determine cluster membership. We obtained the first
ever proper motions for \object{NGC~6253} thanks to the availability
of 4 $V$-band images (Table~\ref{tab:esoarchive}), acquired
with the MPG/ESO WFI on March 30, 2000 as part of the PRE-FLAMES survey
(Momany et al. 2001).  Thus, we have a 4.22 yr temporal baseline
considering that our MPG/ESO WFI data were collected in June of 2004.
For the second epoch, we used our six images
acquired in $V$ filter and listed in Table~\ref{tab:LaSillaObsbvi}.

\begin{table}
\caption{
First-epoch observations taken from the ESO WFI archive.
\label{tab:esoarchive}
}

\begin{center}
\begin{tabular}{c c c c}
\hline
 Date & Filter & Exp (s) & N.imm \\
\hline
 30/3/2000  &  $V$  &   30   & 2 \\
 30/3/2000  &  $V$  &  240   & 2 \\
\hline
\end{tabular}
\end{center}
\end{table}

We used the technique described in Anderson et al.~(\cite{anderson06}), which
allows us to quantify the WFI geometric distortions. One of the key features
of this technique is the so-called local transformation approach for
calculating proper motions. A representative set of local reference
stars can be selected around each target. It is highly desirable that the
cluster members themselves form this set, because the expected internal
velocity dispersion for clusters stars is negligible with respect
to the measurement errors. Assuming that the internal velocity dispersion
is about 1 km~s$^{-1}$, at the distance of \object{NGC~6253}, $d=1.5$~kpc,
it translates into $\epsilon_{\mu}=0.14\,$mas~yr$^{-1}$, which is by
a factor of 10 less than the accuracy of our proper motions.
Once the pixel coordinates of stars have been corrected for geometric
distortions, we selected 50 probable cluster members (with isolated and
unsaturated images) around each target. These stars were used a reference
frame to calculate a linear transformation of coordinates between the
two CCD frames taken at different epochs, which constitute a pair of
``plates". The local transformation effectively takes out a possible
small local residual distortion
pattern.  In general, this transformation is performed iteratively,
starting with photometrically selected cluster members. In each following
step the likely proper-motion nonmembers are deleted and the calculations
are repeated using a cleaner list of cluster members.

\begin{figure}[!]
\center
\includegraphics[width=5cm]{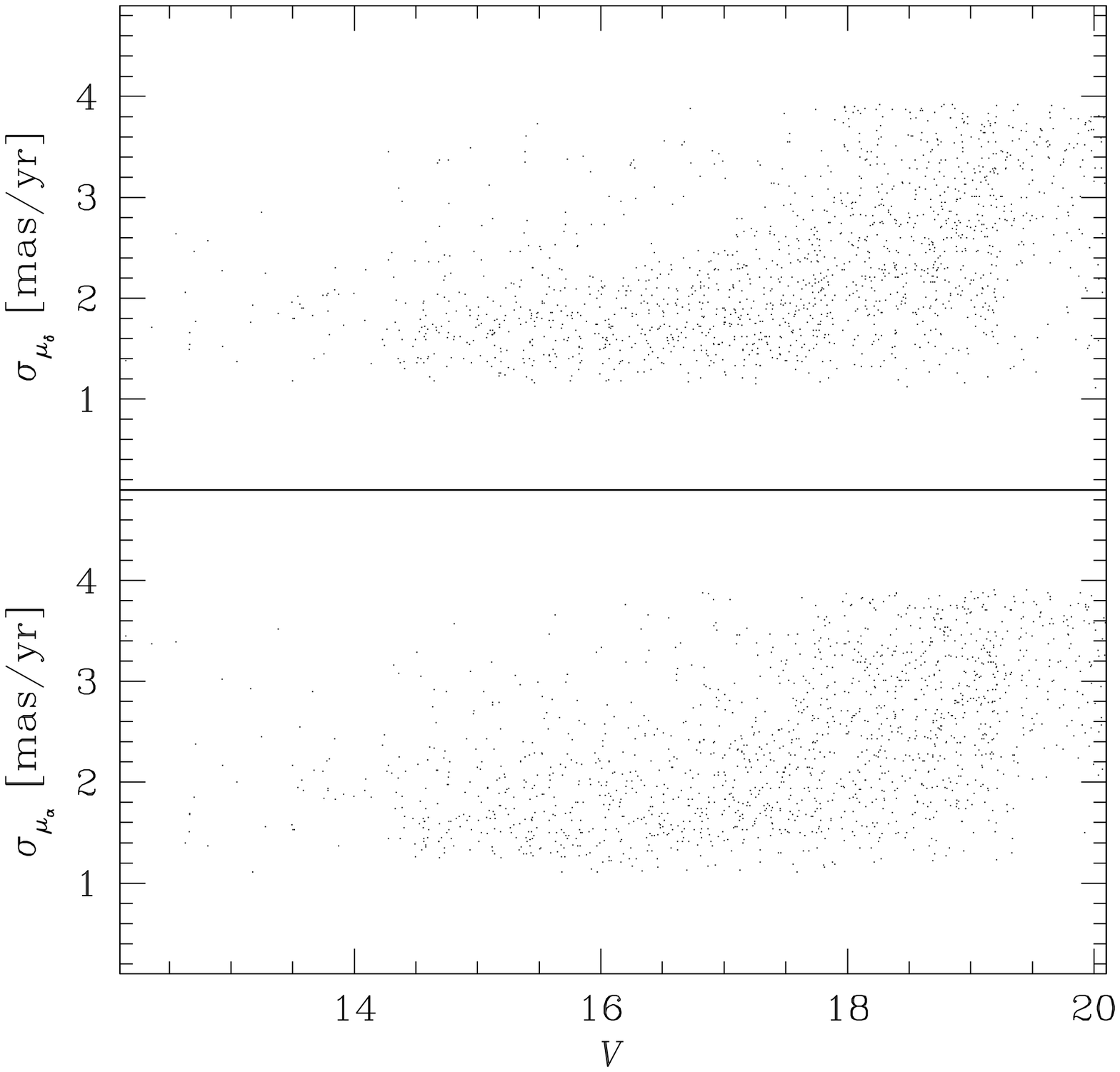}
\includegraphics[width=5cm]{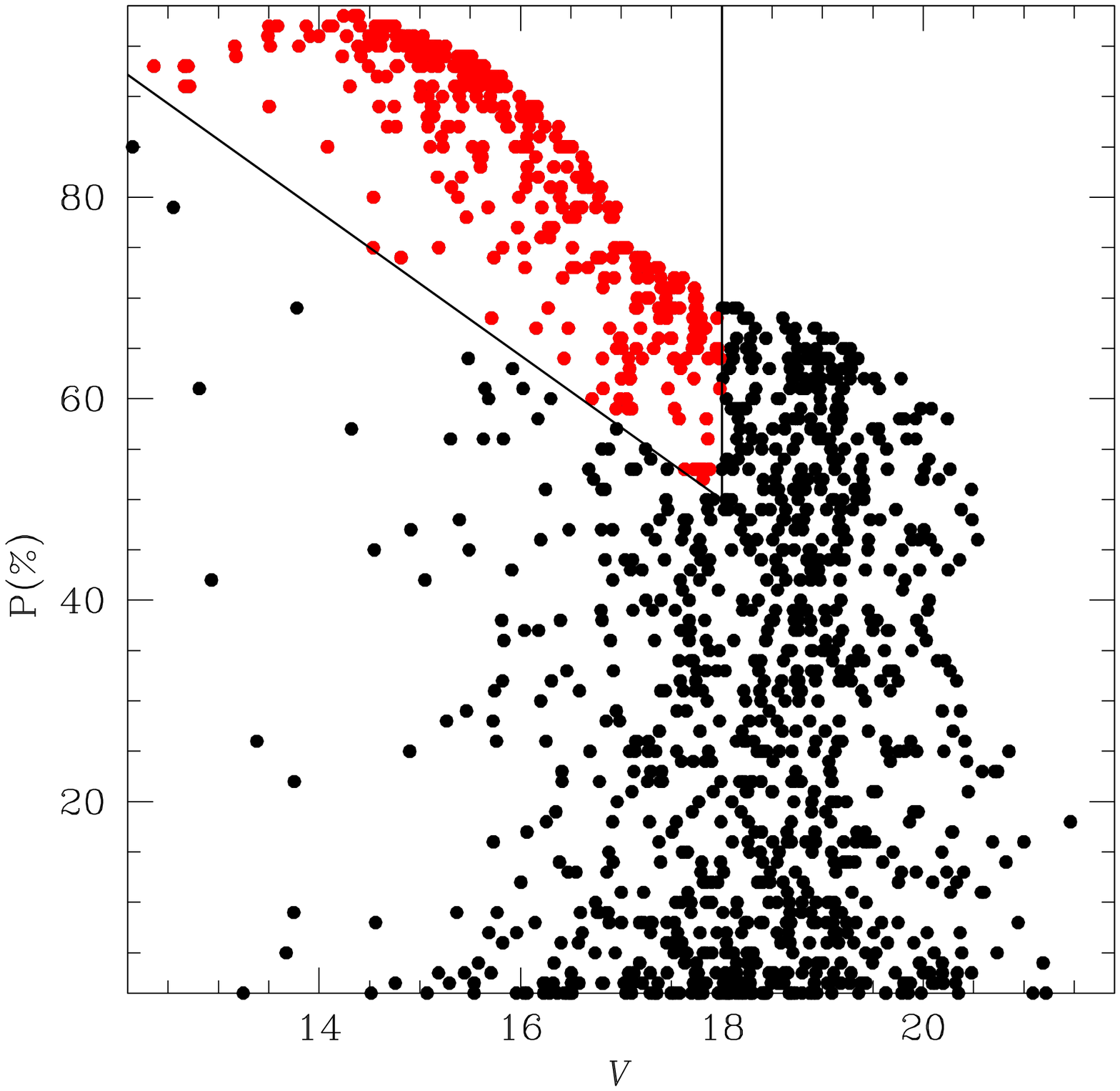}
    \caption{Upper panel: proper motion errors in $\mu_{\alpha}$
and $\mu_{\delta}$ as a function of $V$ magnitude.
Lower panel: membership probability plotted against $V$ magnitude.
The selected probable cluster members (red points) are  enclosed
between the two solid lines.
             }
\label{fig:pmerr}
\end{figure}

The proper motion errors have been estimated from the $rms$ scatter of
calculated proper motions calculated by considering all pairs of first
and second epoch images. Because of the limited number of images used in
proper motion calculations, the resulting errors generally underestimate
the true scatter. As demonstatrated by Anderson et al.~(2006), for
well-exposed images taken with MPG/ESO WFI imager the positional accuracy
of a single image is about 7~mas in each coordinate. Formally, this
translates into the proper motion error of 1.1 mas~yr$^{-1}$ over
the 4.22 yr span. In order to provide a more realistic estimate of
proper motion errors, we added this 1.1 mas~yr$^{-1}$ to the error
estimate provided by the pipeline. As expected, the adjusted proper motions
uncertainties tend to increase with magnitude and reach $2\,$mas yr$^{-1}$
at $V\,=\,18$.  We note that the final errors shown in Fig.~\ref{fig:pmerr}
have an upper limit cut-off at $\sim$4~mas~yr$^{-1}$. The proper motions
with accuracies lower than this limit are not considered in this study.

\begin{figure}[!]
\center
\includegraphics[width=9cm]{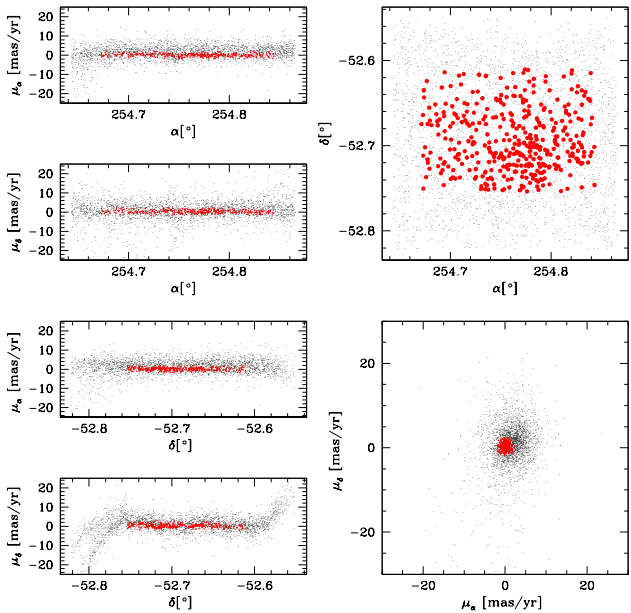}
    \caption{{\rm Left panels}: spatial distributions of proper
      motions along equatorial coordinates $\alpha$ and $\delta$;
      {\rm upper right panel}: spatial distribution of stars in the
      cluster region (chip CCD~51 in the MPG/ESO WFI imager);
      {\rm lower right panel}: proper motion diagram.
      The stars which have passed our selection criteria for a probable
cluster member are indicated by the red dots.
      }
\label{fig:selxy}
\end{figure}

We note that in the highlighted range of spatial distribution
(Fig.~\ref{fig:selxy}) our proper motions are free of potential spatial
correlations.  Outside this range, we were unable to adequately correct
for the geometric distortion effects. This is likely due to the fact that
the cluster is essentially contained by the inner CCD No.~51 of the
MPG/ESO WFI imager. The local transformation method relies heavily on
cluster members which are rare in the outer parts of a cluster.
As the result, the calculated proper motions may get biased
as we move outwards from the cluster center.
In order to isolate reliable cluster's members, we consider only
the stars whose coordinates are comprised between 200$<$x$<$1800 pixels
and 1000$<$y$<$3000 pixels on the system of CCD No.~51 which contains
most of the \object{NGC~6253} members.Nevertheless, in the final catalog we
retain all proper motions we have calculated.

\subsection{Cluster membership}

\begin{figure*}
\center
\includegraphics[width=15cm]{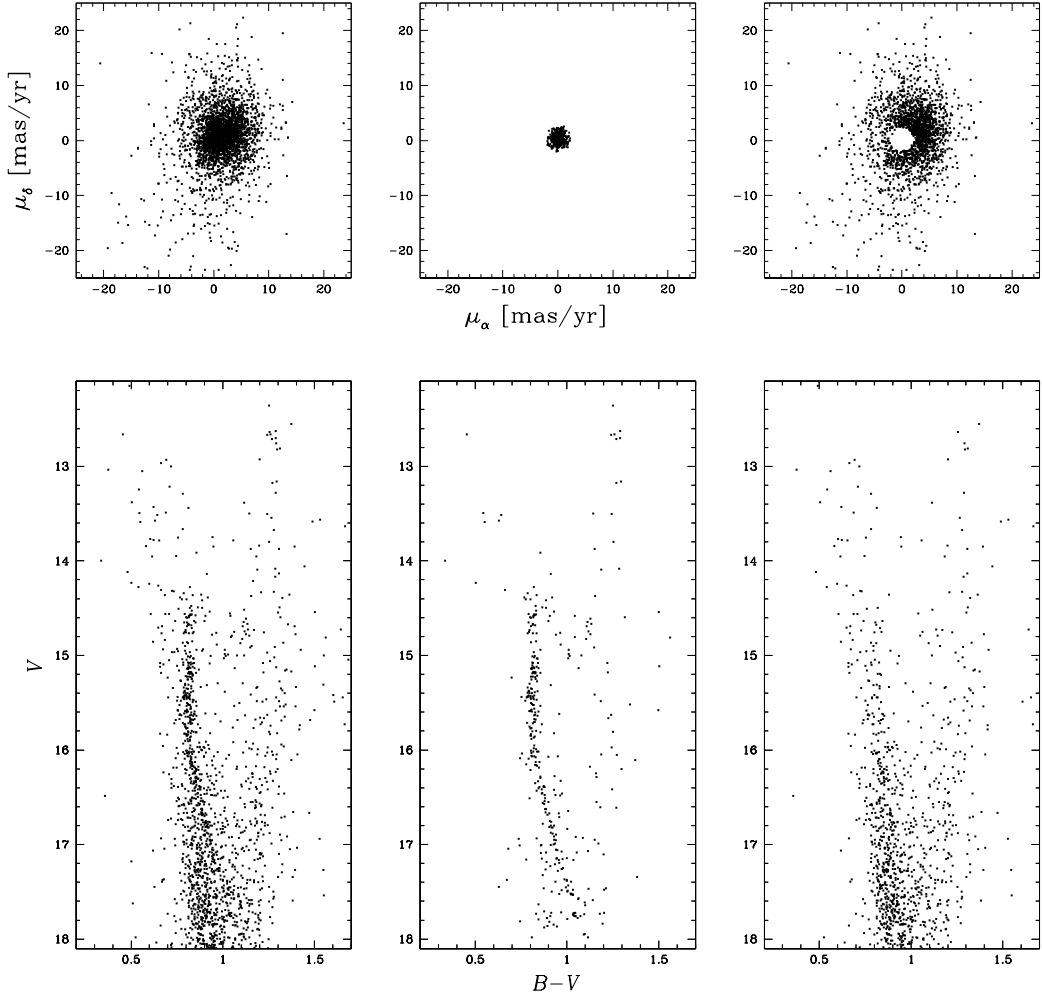}
    \caption{Proper motion vector-point diagrams (upper panels) along
with correspondent CMD diagrams (lower panels). Left panels: all stars
brighter than $V=18.1$ are considered; middle panels: only the selected
cluster members; right panels: mostly field stars.}
\label{fig:selection}
\end{figure*}

As      illustrated    in      Fig.~\ref{fig:selection}, the proper motion distribution shows
the      presence      of     two   distinct populations of stars: one  has  a narrow peak at
$\mu_{\alpha}\,=\,\mu_{\delta}\,=\,0$ mas~yr$^{-1}$  and represents the cluster's stars, the other one
represents          the        broader         distribution     of       field         stars.
Following    the    approach    proposed by Vasilevskis (1958) the   proper-motion membership
probability  can be expressed as:

\begin{equation}
\rm P_\mu=\frac{\Phi_c}{\Phi_c+\Phi_f},
\end{equation}

where     $\rm \Phi_c$    is    the   distribution    of  cluster  stars,  and  $\rm \Phi_f$   the
distribution  of  field  stars  in the proper motion diagram, which are generally represented by
Gaussian      functions.        As        in           Yadav      et       al.~(2008),    the
calculation of the membership probabilities was done following the local sample method, where
for      each target star a surrounding sample of stars is selected to closely represent the
properties of the target. The local sample stars were selected inside a 2.5 mag range centered
on the target which allowed us to compensate for the effect of a magnitude-dependent cluster-to-field
star ratio. We parameterized the proper motion dispersion for both,
the cluster and field stars, and the center of field star
distribution as a function of the magnitude.
The center of cluster star distribution was always kept fixed at
$\mu_{\alpha}\,=\,\mu_{\delta}\,=\,0$ mas~yr$^{-1}$.
Figure \ref{fig:pmerr} (lower panel) shows the magnitude dependence of the membership probabilities,
which reflects the degree of reliability of our proper motions.
Towards  fainter  magnitudes,  the
increasing   level   of    field  star contamination is resulting
in a steadily decreasing maximum membership probability.
 The separation between cluster and field is convincing for $V\,<\,18$,
for which the cluster main sequence can easily be distinguished.
We thus adopted a magnitude dependent threshold to distinguish between
the cluster and field stars. We require the probable
cluster members   to       have their membership  probabilities $>$90\%
at $V\,=\,12.5$, and $>$50\% at $V\,=18$, assuming a linear dependence of the
threshold between these two extremes. We found a total of $383$ stars
that satisfy the criteria reported above. Nevertheless as discussed in
Sec.~\ref{s:twocolors} and in Sec.~\ref{s:binaries} some residual
contaminants are certainly present in this sample, in particular field stars that 
share the proper motion of cluster stars.

\subsection{Membership of spectroscopic stars}
\label{s:spectroscopy}

There are two spectroscopic studies of \object{NGC~6253} that list
individual stars belonging to the upper main sequence or the red giant
branch: Sestito et al. (2007, SE) and  Carretta, Bragaglia \& Gratton
(2007, CA2).  Thus, SE derived radial velocities for seven stars 
stretching from the main sequence turnoff to the red clump of \object{NGC~6253}, using
high-resolution spectroscopy at VLT/UVES.
Among these stars, they isolated four probable members with the
mean radial velocity of $-29.71\,\pm\,0.79$ km~s$^{-1}$. The other two stars,
069360 and 022182 enumerated by Momany at al. (2001), appear to be
nonmembers according to their radial velocities Finally, star 023501
is classified by SE as a doubtful member. In Table~\ref{tab:rvpm}, 
we show our membership probabilities
for this sample of SE stars. First column reports for each star
the number ID$_{\rm EIS}$ taken from the EIS survey (Momany et al. 2001).
Second column shows our identifier ID$_{\rm MO}$. The following RA, DEC,
$V$ and $B-V$ are all from our catalog, while radial velocities along
with their errors are from SE. The next column lists proper-motion
membership probability (P). Metallicity [Fe/H] and its error is from
Table 4 in SE. The last column shows the status of a star --
TO: turn-off star; RGB: red giant branch star; SGB: sub-giant branch star;
clump: red clump star; M: probable cluster's member; NM: probable non member;
bin?: probable binary star. We confirm that star 022182 is a field
star, which also explains its lower metallicity ([Fe/H]=$+$0.12).
For all other stars we obtained membership probabilities larger
than P=$93$\%. A deviant radial velocity for star 069360 is more
suggestive of its binarity and cluster membership than the likelihood of
being a field star. The same conclusion can be drawn for star 023501.

The study by CA2 (Table~\ref{tab:rvpm1}) analyzed high resolution spectra of five red clump stars
in \object{NGC~6253} obtained with the UVES and FEROS spectrographs.
According to our proper motions, all five stars have high membership
probabilities and are considered to be the cluster members. Star
2508 (in BR enumeration) is very likely to be a cluster member and
a spectroscopic binary as indicated by its deviant radial velocity
listed by CA2. 

\begin{table*}
\caption{
Proper-motion membership probabilities for stars listed in the SE paper.
\label{tab:rvpm}
}

\begin{center}
\begin{tabular}{c c c c c c c c c c}
\hline
 ID$_{\rm EIS}$ & ID$_{\rm MO}$ & RA & DEC & $V$ & $B-V$ & RV$\,\pm\,$rms & P & [Fe/H]$\,\pm\,$err & Status\\
            &           & (J2000) & (J2000) & & & (km s$^{-1}$) & \% &\\
\hline
 069885 & 45485 & 254.717971 & -52.694686 & 14.341 & 0.786 & -28.81$\pm$1.30 & 98 & +0.45$\pm0.11$ & TO, M \\
 023501 & 45474 & 254.721883 & -52.712070 & 14.275 & 0.819 & -33.49$\pm$2.87 & 96 & +0.29$\pm0.11$ & TO, M, bin? \\
 069360 & 45421 & 254.779357 & -52.706134 & 13.175 & 1.269 & -21.07$\pm$0.59 & 94 & +0.48$\pm0.12$ & RGB, M, bin? \\
 022182 & 45398 & 254.814677 & -52.725906 & 13.840 & 1.199 & -35.51$\pm$0.77 &  0 & +0.12$\pm0.09$ & SGB/RGB, NM \\
 023498 & 45455 & 254.839016 & -52.712035 & 13.912 & 0.855 & -30.13$\pm$1.01 & 96 & +0.32$\pm0.11$ & SGB, M \\
 024707 & 45404 & 254.803860 & -52.693301 & 13.501 & 1.144 & -30.58$\pm$0.65 & 97 & +0.39$\pm0.11$ & SGB/RGB, M \\
 105495 & 45415 & 254.713319 & -52.616005 & 12.700 & 1.289 & -29.33$\pm$0.84 & 93 & +0.49$\pm0.13$ & clump, M \\
\hline
\end{tabular}
\end{center}
\end{table*}

\begin{table*}
\caption{
The same as in Table~\ref{tab:rvpm} but for the CA2 stars. The identifier
ID$_{\rm BR}$ is from Bragaglia et al. (1997).
\label{tab:rvpm1}
}

\begin{center}
\begin{tabular}{c c c c c c c c c c}
\hline
 ID$_{\rm BR}$ & ID$_{MO}$ & RA & DEC & $V$ & $B-V$ & RV & P & [Fe/H]$\,\pm\,$err & Status\\
            &           & (J2000) & (J2000) & & & (km s$^{-1}$) & \% &\\
\hline
2509 &	  45412	&  254.816479  &  -52.707251 &   12.664  &   1.259 & -28.71 & 93 & +0.45$\pm$0.11 & clump, M\\
2885 &	  45410 &  254.722210  &  -52.698323 &   12.629  &   1.289 & -28.13 & 91 & +0.47$\pm$0.14 & clump, M\\
4510 &	  45414 &  254.775449  &  -52.665473 &   12.713  &   1.268 & -27.44 & 91 & +0.42$\pm$0.12 & clump, M\\
2508 &	  45413 &  254.766037  &  -52.707004 &   12.666  &   1.241 & -20.60 & 91 &       -        & clump, M, bin?\\
3595 &	  45407 &  254.774389  &  -52.684524 &   12.360  &   1.252 & -28.76 & 93 & +0.48$\pm$0.09 & clump, M\\
\hline
\end{tabular}
\end{center}
\end{table*}


\begin{figure}
\center
\includegraphics[width=8cm]{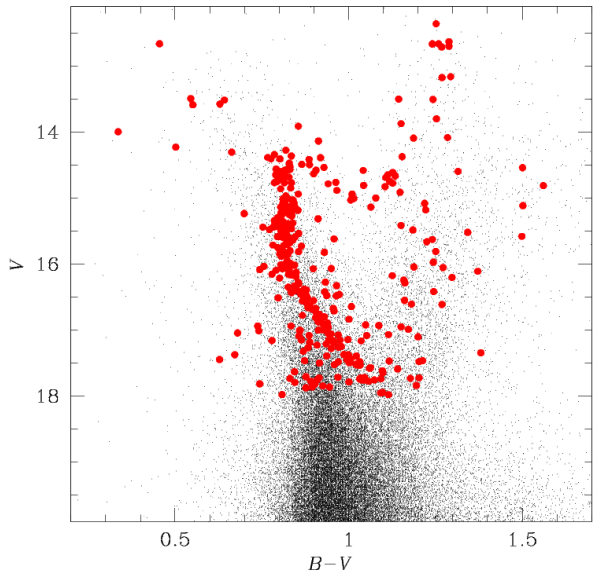}
    \caption{A $BV$ CMD for the entire field-of-view, imaged by the MPG/ESO
 WFI. The stars meeting our cluster membership
        selection criteria (Sect.~\ref{s:propermotion}) are highlighted by the
        red dots.
        }
\label{fig:cmdtot}
\end{figure}

\section{Analysis of the color-magnitude diagrams}
\label{s:cmd}
\begin{figure*}
\center
\includegraphics[width=15cm]{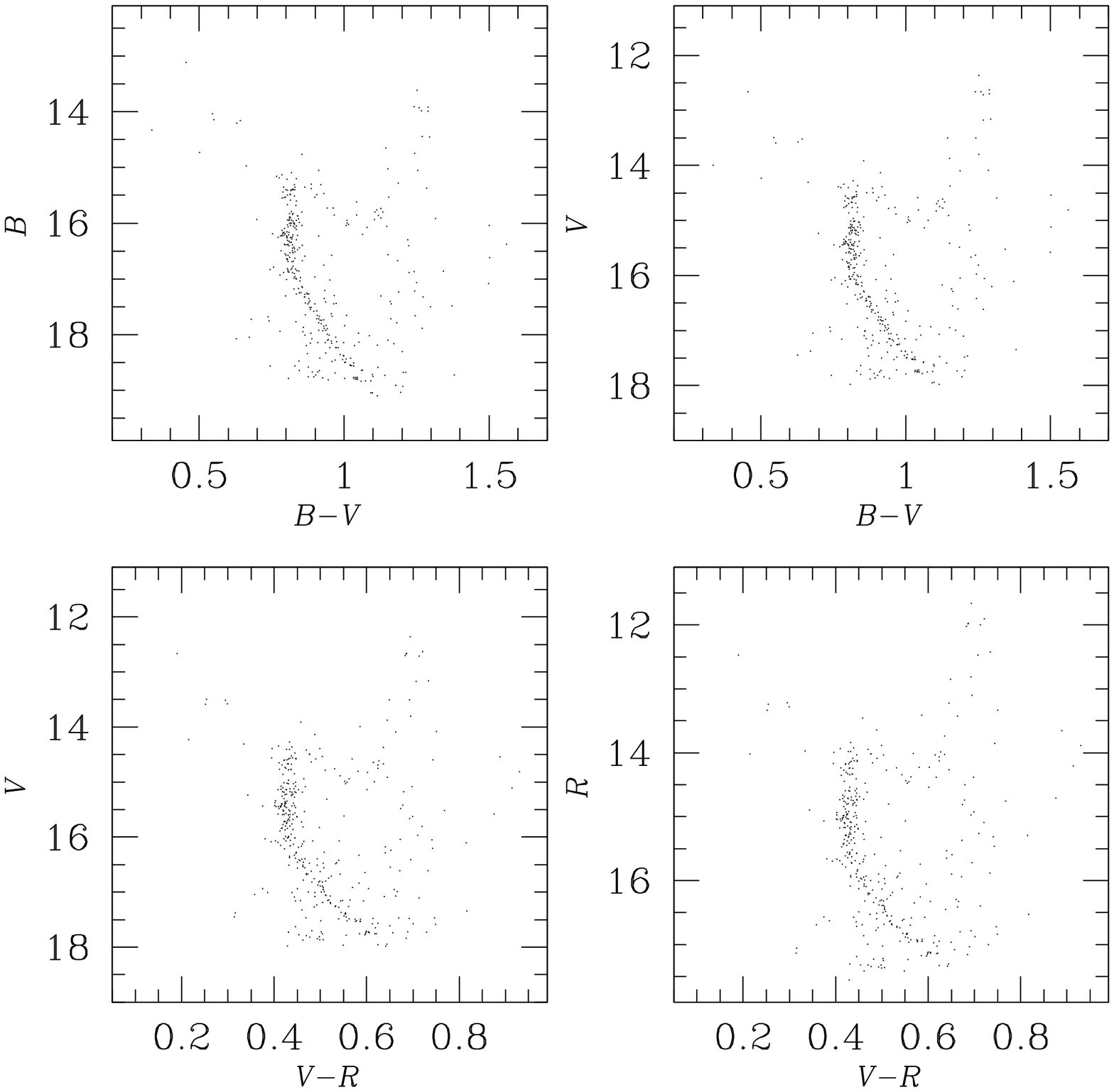}
    \caption{Color magnitude diagrams for the proper-motion selected
cluster stars.
        }
\label{fig:cmd}
\end{figure*}

The observed color-magnitude diagrams of~\object{NGC~6253} in various
optical/near-infrared bandpasses are presented
in Fig.~\ref{fig:cmd}.  Only those stars meeting our cluster member selection
criteria, defined in Sect.~\ref{s:propermotion}, are plotted in these CMDs.
Among various features of CMDs, one of them is the hook at the turnoff
of main sequence already noted by BR. In addition, sub-giants, red giant branch
stars,  red clump, and blue stragglers are all present. We note that
due to the relatively high uncertainties of our near infrared photometry
the appearance of cluster sequences and populations is less prominent.

\begin{figure*}
\center
\includegraphics[width=8cm]{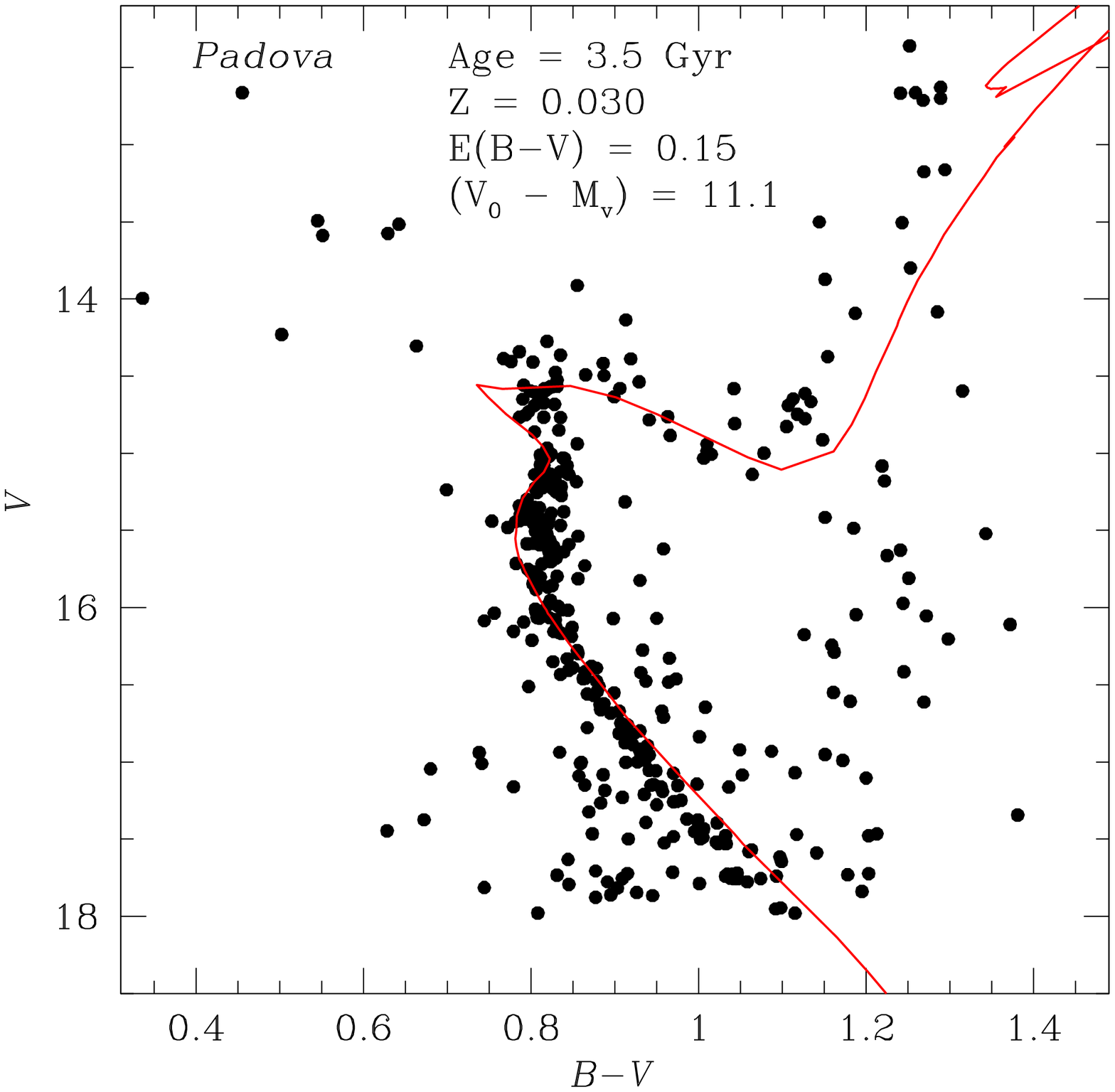}
\includegraphics[width=8cm]{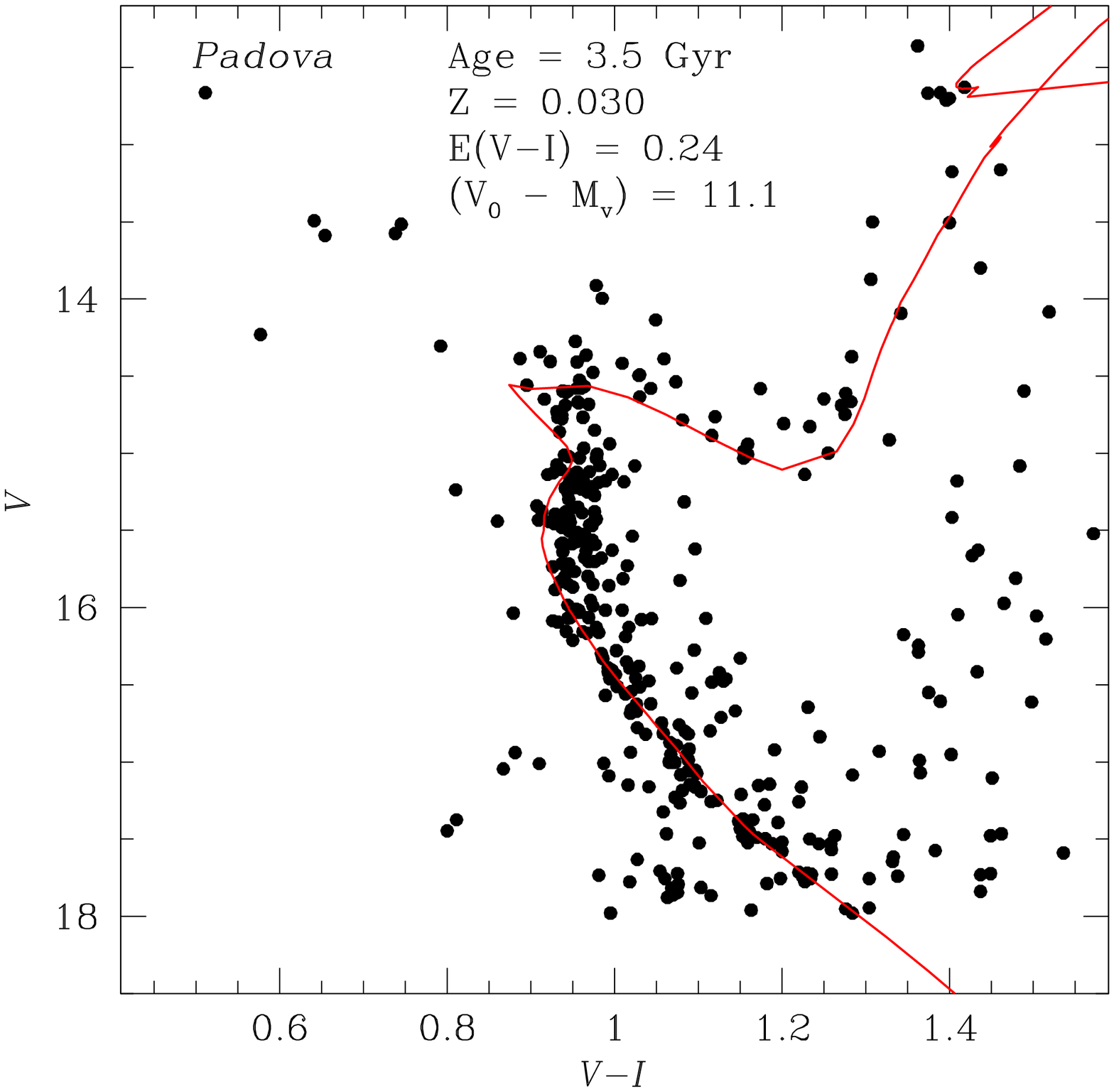}
\includegraphics[width=8cm]{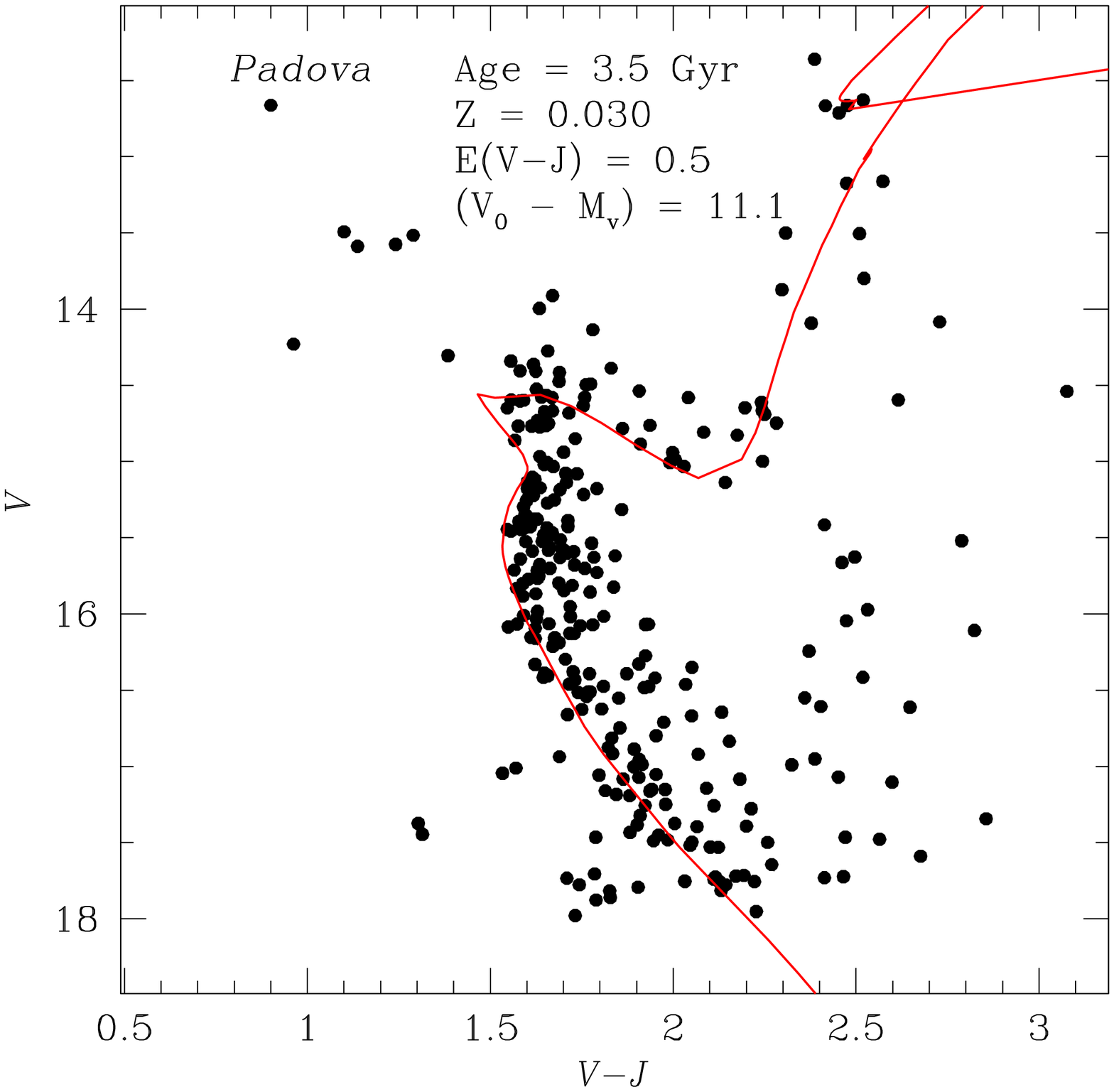}
\includegraphics[width=8cm]{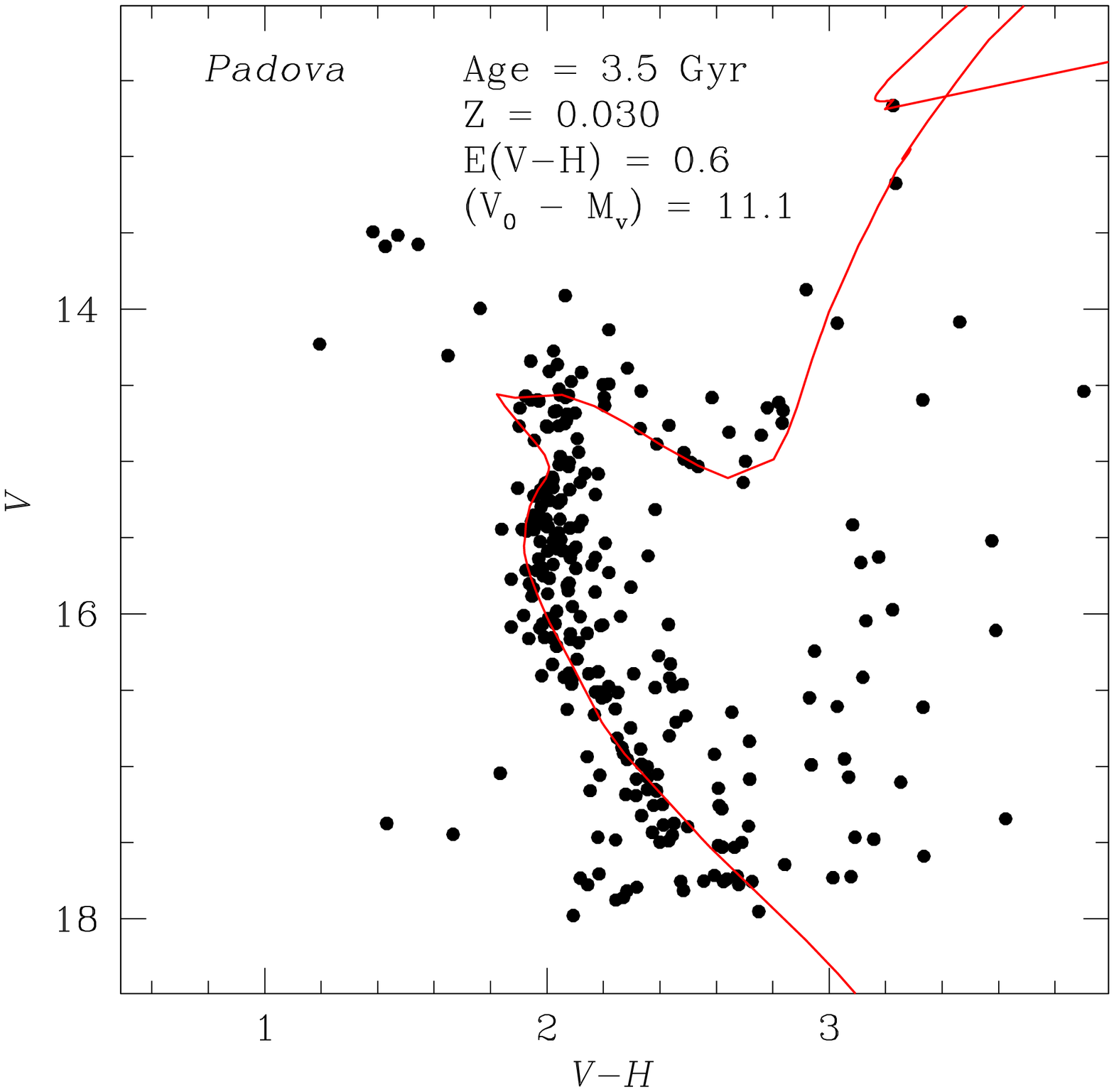}
    \caption{Best   fit    Padova     stellar    isochrones   superimposed on
              $(V, B-V)$,  $(V, V-I)$,  $(V, V-J)$ and  $(V, V-H)$ CMDs of \object{NGC~6253}.
        }
\label{fig:isochrones_pad}
\end{figure*}

\begin{figure*}
\center
\includegraphics[width=8cm]{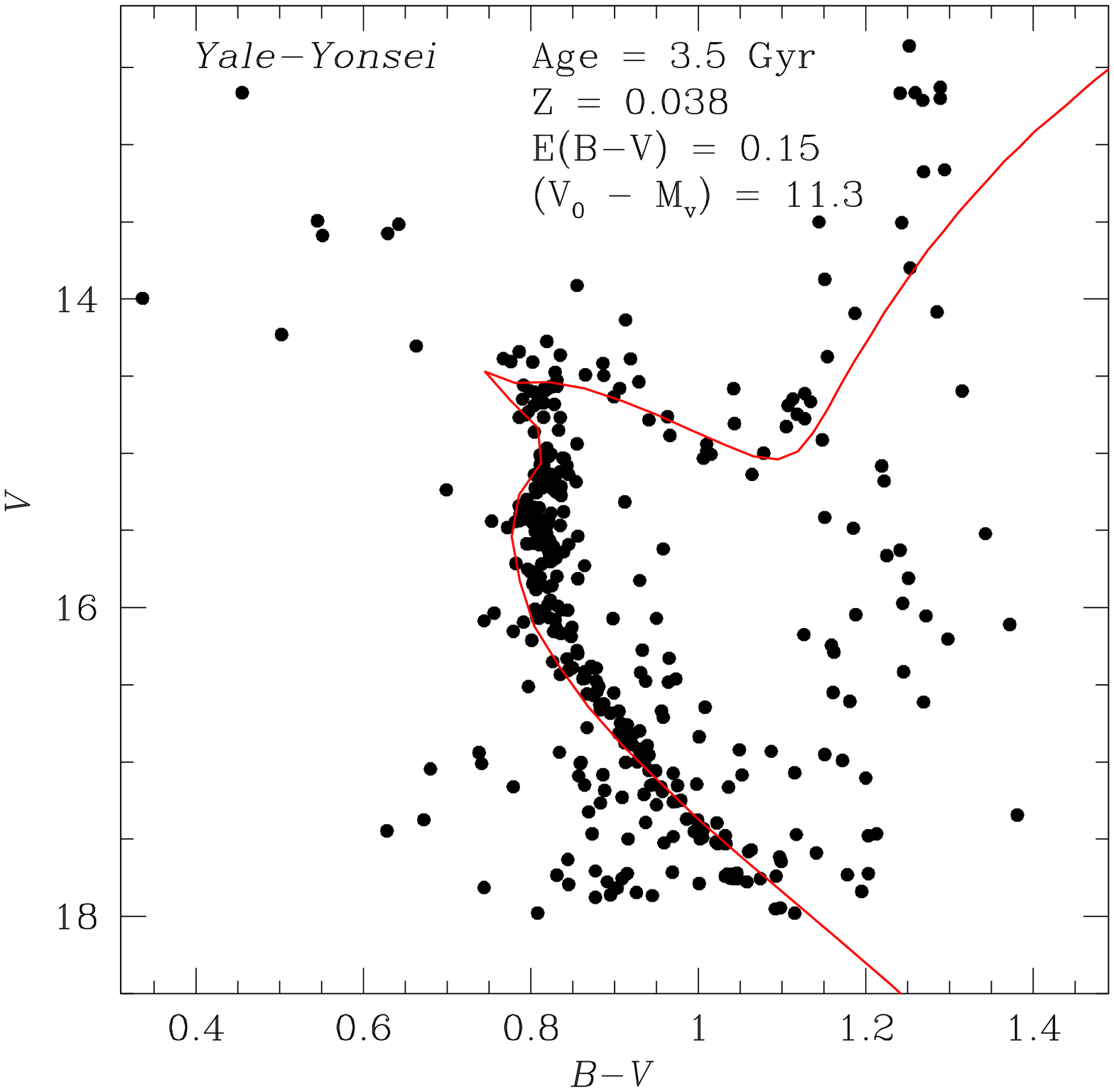}
\includegraphics[width=8cm]{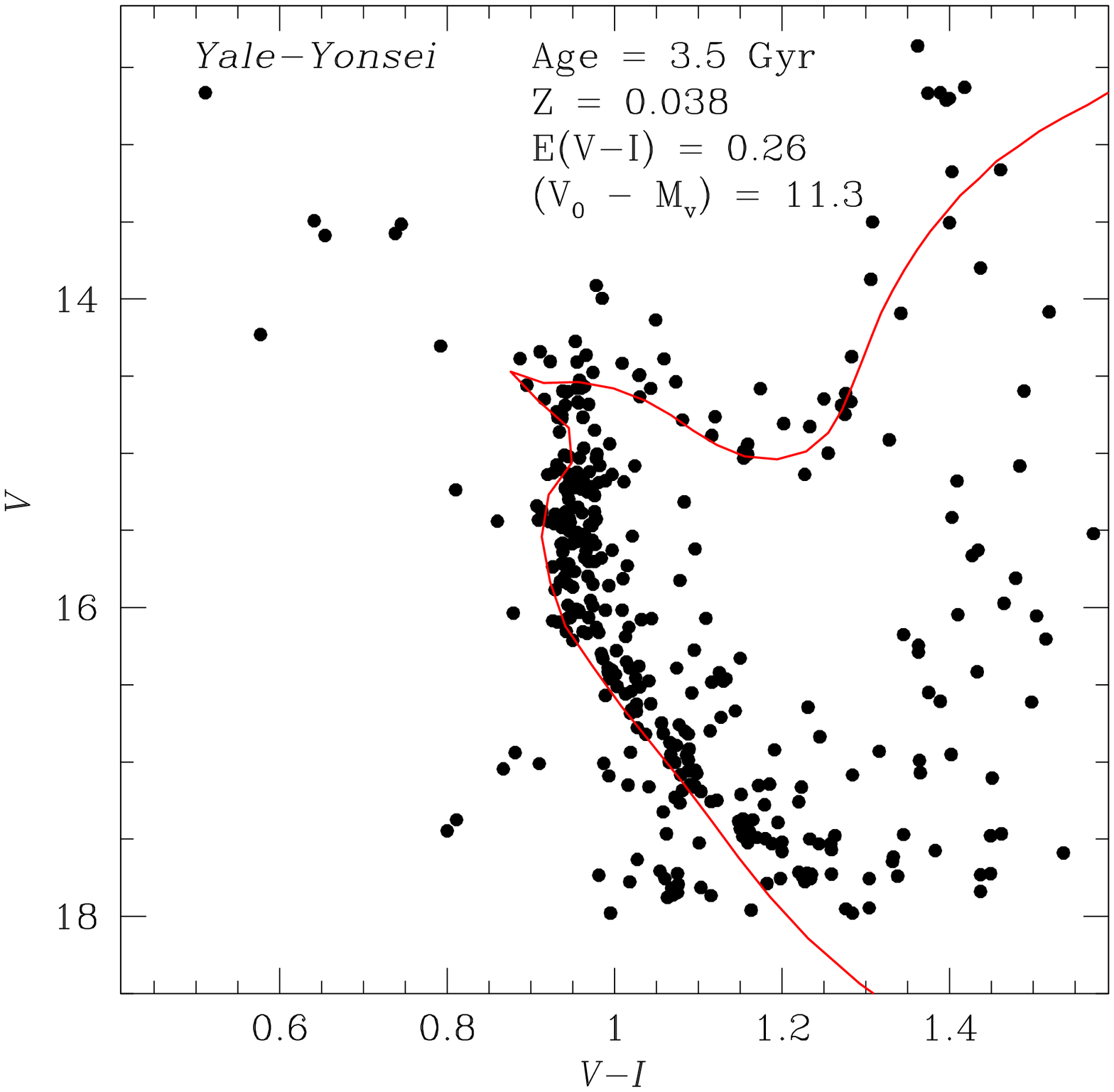}
\includegraphics[width=8cm]{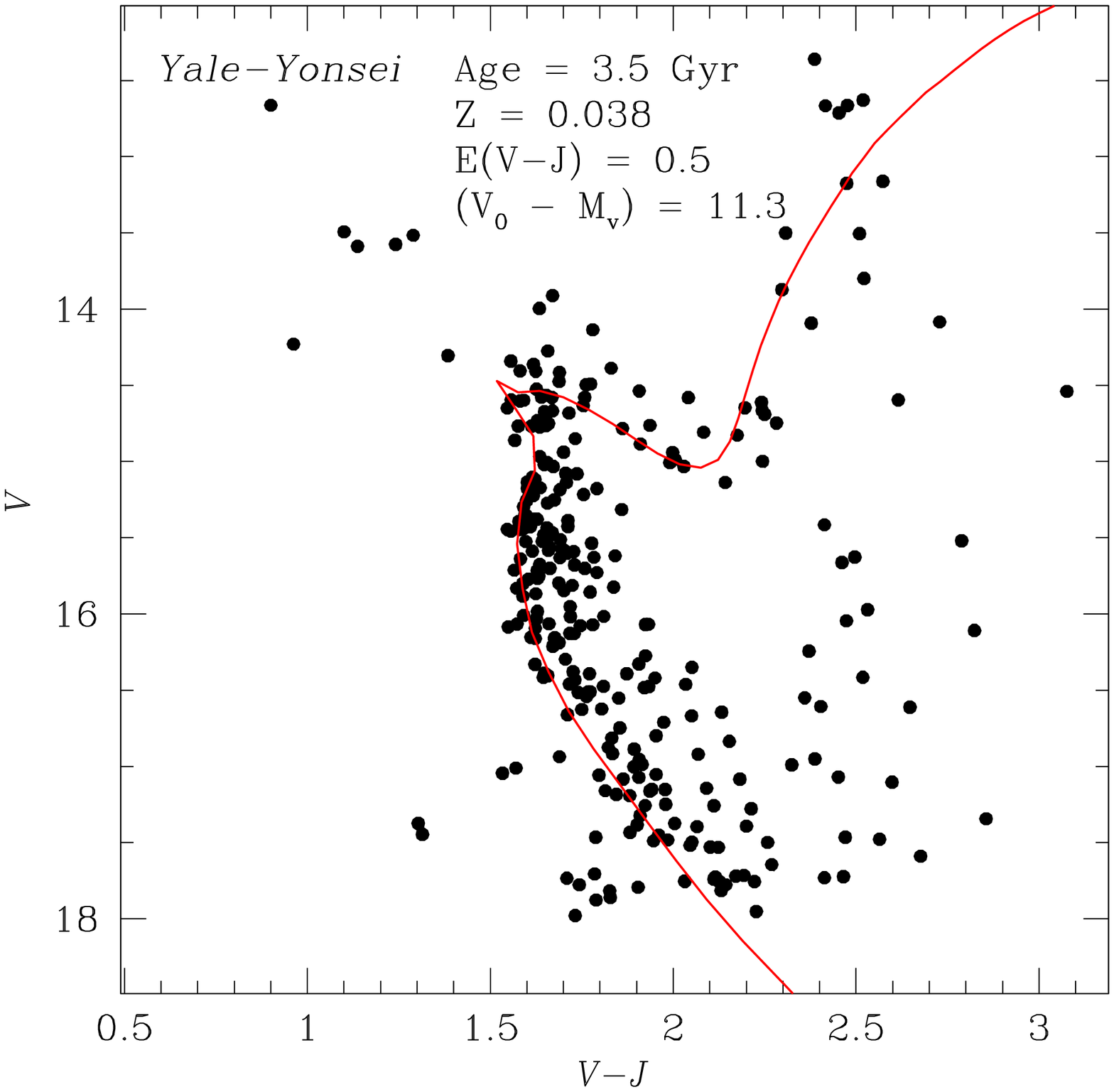}
\includegraphics[width=8cm]{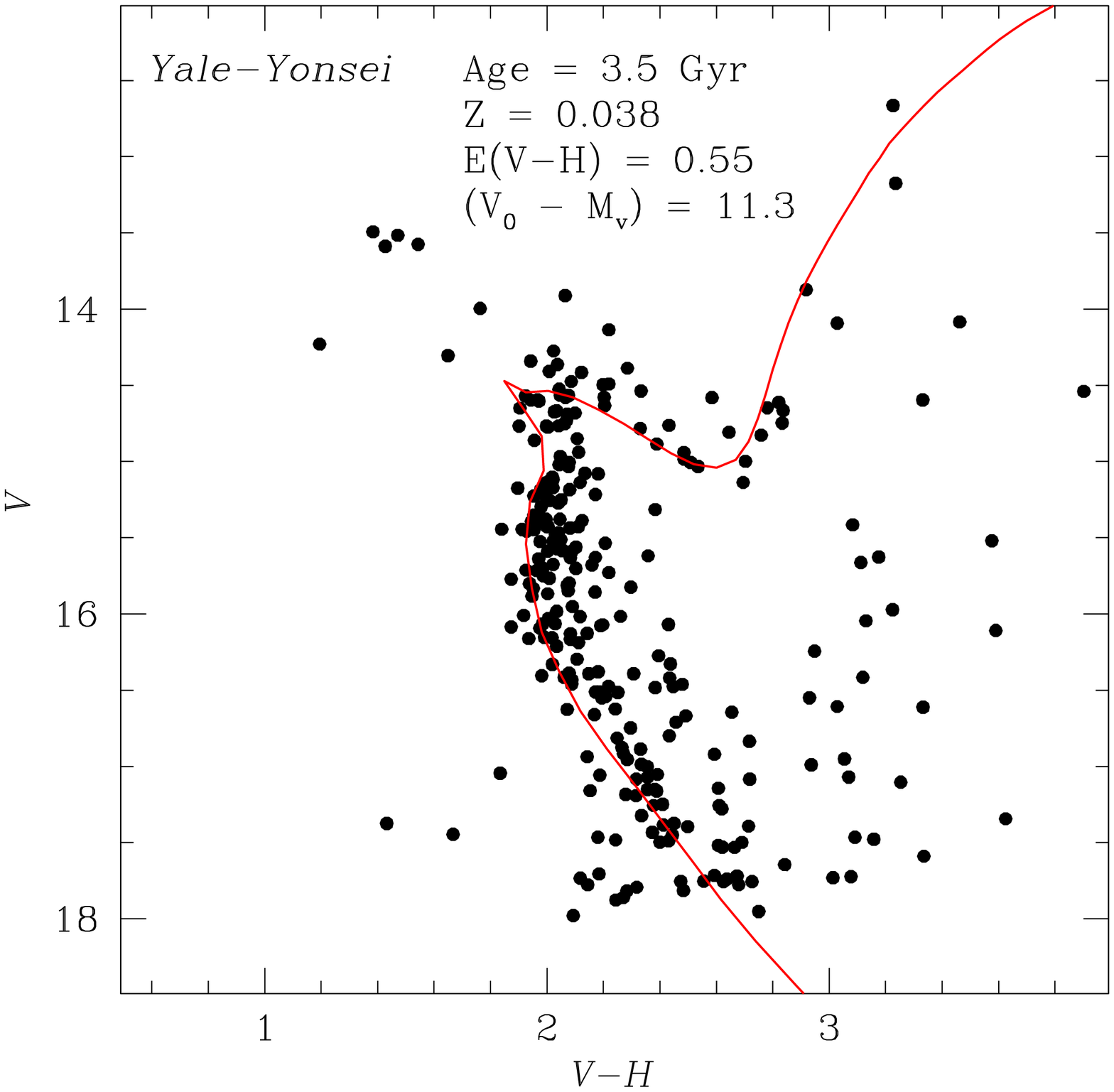}
    \caption{Best   fit    Yale-Yonsei     stellar    isochrones   superimposed on
              $(V, B-V)$,  $(V, V-I)$,  $(V, V-J)$ and  $(V, V-H)$ CMDs of \object{NGC~6253}.
        }
\label{fig:isochrones_yale}
\end{figure*}

We    used        the       Padova    and Yale-Yonsei set of isochrones to fit the
\object{NGC~6253}    CMDs    using    both    optical and near infrared bandpasses. As noted in previous   spectroscopic studies of \object{NGC~6253},
the solar-scaled models are the most appropriate for this open                  cluster.  From      the    Padova    database~\footnote{http://stev.oapd.inaf.it/cgi-bin/cmd}
(Girardi et al.~2002,                Marigo et al.~2008),                       we
obtained                a             set              of solar-scaled  isochrones
with                  metallicity        Z=0.03, which is the highest heavy
metal content for this set of isochrones.           As      shown        in
Fig.\ref{fig:isochrones_pad},  the  best  fit is obtained for the  age
$\sim$3.5~Gyr.  We fixed the true distance modulus in the $V$ bandpass
  at  $(V_0-M_v)\,=\,11.1$, adopting
 $\rm E(B-V)\,=\,0.15$, $\rm E(V-I)\,=\,0.24$,  $\rm E(V-J)\,=\,0.5$ and
 $\rm E(V-H)\,=\,0.6$                                                          from
the simultaneous best fit to the $(V, B-V)$, $(V, V-I)$, $(V, V-J)$ and $(V, V-H)$ CMDs.
In         the       same      manner,   from
the    Yale       database~\footnote{http://www.astro.yale.edu/demarque/yyiso.html}
(Demarque et al. 2004),  we obtained a set of solar-scaled isochrones with Z=0.038 (equivalent to [Fe/H]=+0.36).      Again,   the  best  fit (Fig.~\ref{fig:isochrones_yale})  is
obtained               for  the      age $\sim$3.5~Gyr. Similarly to the fit
with Padova isochrones, here we obtain
$(V_0-M_v)\,=\,11.3$,        $\rm E(B-V)\,=\,0.15$,         $\rm E(V-I)\,=\,0.26$,
$\rm E(V-J)\,=\,0.5$                    and                 $\rm E(V-H)\,=\,0.55$.
We note that the values of ratios  $\rm E(V-I)/E(B-V)=1.6,1.7$          and
 $\rm E(V-H)/E(V-J)=1.2,1.1$ obtained from the fit with the Padova and
Yale isochrones, respectively, to the appropriate CMDs is consistent with the values reported by
Mathis (1990) for the standard reddening law ($\rm E(V-I)/E(B-V)=1.6, E(V-H)/E(V-J)=1.2$).
Thus, we           confirm the result     of      SA that
the    color excess ratios are in agreement with the assumption
of  a normal reddening law in the direction of \object{NGC~6253}.
The fundamental parameters we have obtained  are consistent with those
derived by PI,  SA, AT within the quoted uncertainties. As indicated in
 Table~\ref{tab:results}, BR    have       derived          a       higher
reddening   in         $\rm E(B-V)$  and, reciprocally, a smaller distance modulus. Also,   TW   indicated a higher amount of reddening, but   this   estimate  was subsequently
revised   in the study of AT, resulting in $\rm E(B-V)\,=\,0.16$. Despite      the small discrepancies  between      the          various                photometries
noted       in      Table~\ref{tab:confphot}, our study confirms a  lower value
for the reddening in $\rm E(B-V)$ and a higher value for the distance modulus
with respect to the values of BR.  The best isochrone-fit implies the age of $3.5$ Gyr,
which is  somewhat  older  than generally believed in previous studies such
as by  BR, SA, and TW.

Regarding                the         overall   isochrone fit to the CMDs
 shown in Figs.~\ref{fig:isochrones_pad}, \ref{fig:isochrones_yale},
we observe that the turn-off morphology is well-reproduced by both       sets    of
isochrones, while the Padova isochrone  provides less accurate fit to the main
sequence around $V\sim18$ as opposed to the Yale isochrone.
The red giant branch is not very well fit by both sets of isochrones,
 especially in the $(V,B-V)$ CMD, where the theoretical isochrone
it is clearly red-shifted  with respect to the observational RGB.
The   red clump location around $V\,\sim\,12.5$ is well-reproduced by the
Padova isochrone,  whereas    the Yale isochrones fail to extend
to this feature because the Yale models do not consider the phase of
helium burning core in stars.

\subsection{The color-color diagram}
\label{s:twocolors}

\begin{figure}
\center
\includegraphics[width=8cm]{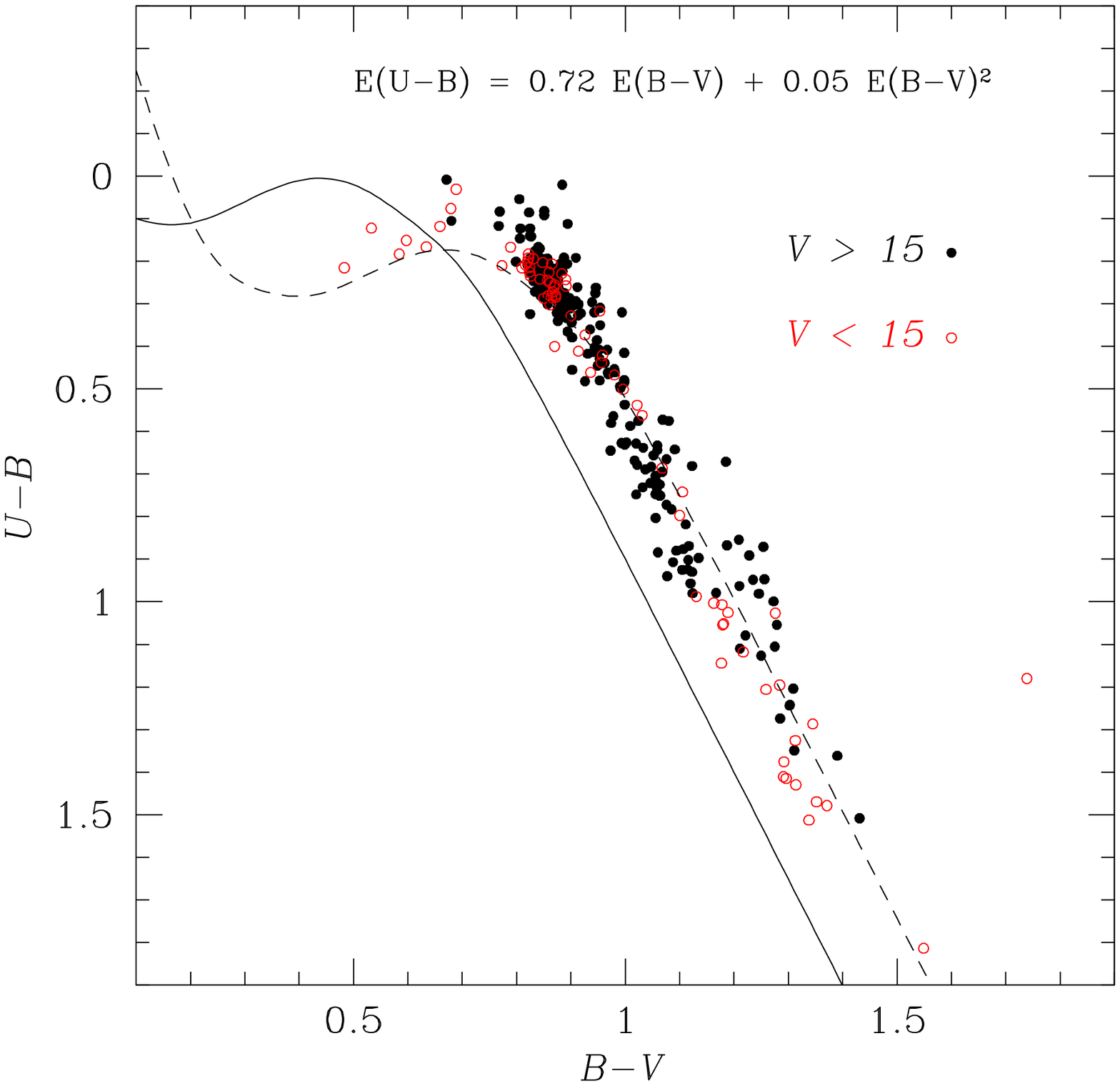}
\includegraphics[width=8cm]{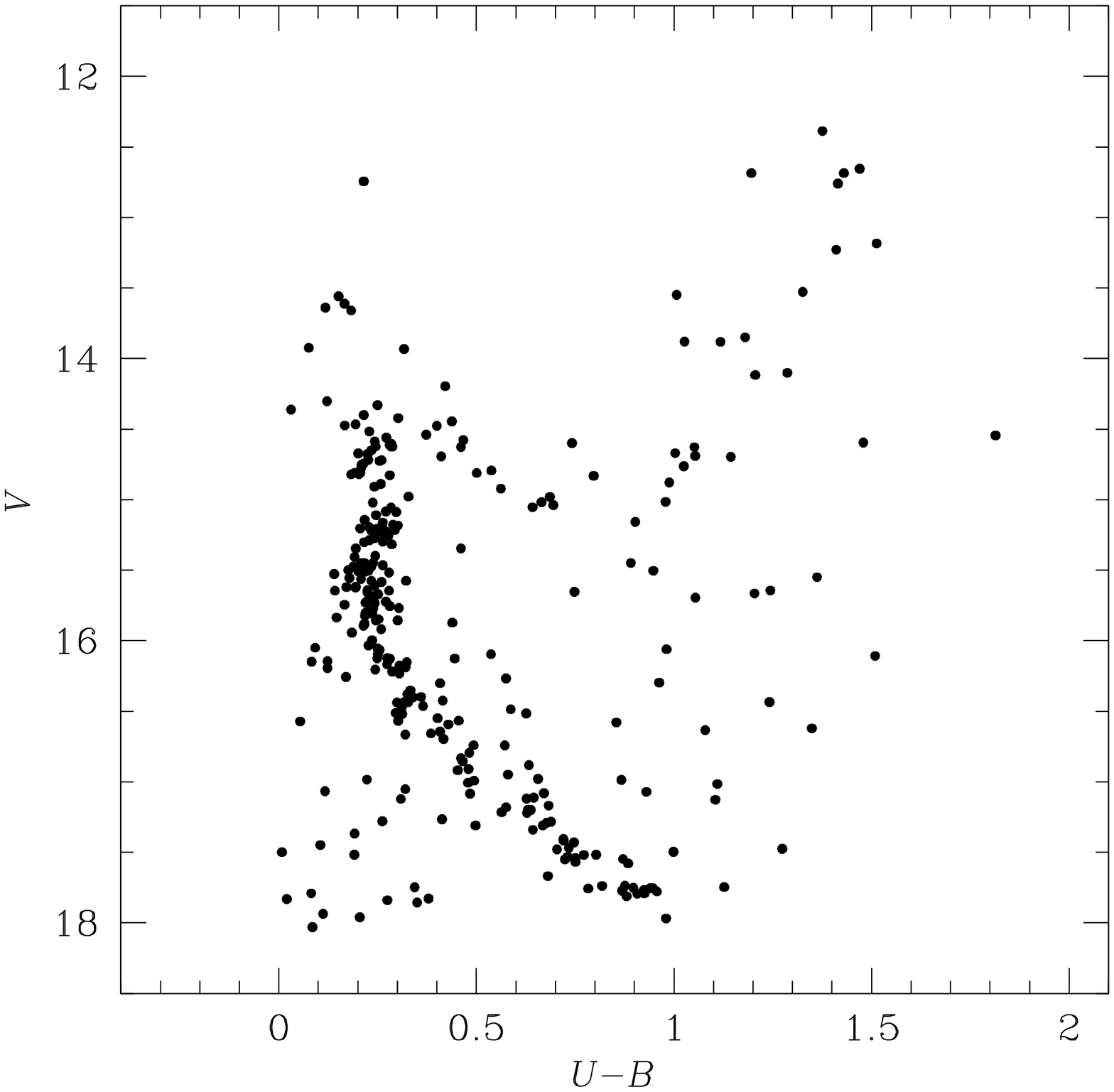}
    \caption{
    {\it Upper panel:} Color-color diagram for \object{NGC~6253}. The likely cluster members are
    selected according to the criteria presented in Sec.~~\ref{s:propermotion}. The solid
    line represents the Schmidt-Kaler~(1982) empirical ZAMS, which is then
    shifted by $E(B-V)\,=\,0.23$ (as obtained by Bragaglia et al.~1997) along the reddening vector
    assuming a normal reddening law, indicated by the dashed line. Open circles (red color) 
    represent stars with $V<15$ and filled circles (black color) stars with $V>15$. 
    {\it Lower panel:} $(V,U-B)$ CMD
    considering all stars from the upper panel. Photometry
    is from Bragaglia et al.~(1997) as our catalog does not include the $U$ band.
    }
\label{fig:bvub}
\end{figure}

Field stars with similar proper motions to those of cluster stars clearly contaminate the
sample of likely cluster members presented in Sec.~\ref{s:propermotion}. This is evident
from the color magnitude diagrams of Fig.~\ref{fig:cmd}. 
In Fig.~\ref{fig:bvub} (upper panel), we show the $(U-B,B-V)$ color-color diagram,
obtained using BR photometry and considering
all stars that meet the cluster membership criteria presented in Sec.~\ref{s:propermotion}.
The solid line represents the Schmidt-Kaler~(1982) empirical ZAMS, and the dashed line
the same ZAMS shifted by $E(B-V)=0.23$ (to be consistent with the BR estimate, see 
Tab.~\ref{tab:results}) along the reddening vector considering a normal standard reddening law.
We cross-correlate our catalog with that one of BR and
found $302$ stars in common satisfying our proper motion membership criteria. 
We use the BR photometry as our catalog does not include the $U$ band.
In Fig.~\ref{fig:bvub} (bottom panel), the stars with $U-B<0.1$ and $V>15$
are clear contaminants as also shown in the two color diagram where these stars tend
to deviate from the ZAMS. On the red side, the distinction is less effective
as binary cluster stars and field stars both have redder colors with respect to the main sequence,
although stars with $B-V>1.4$ are certainly contaminants (Fig.~\ref{fig:cmd}).
Further information
(e.g., radial velocities, abundance analysis) is necessary to clearly disentangle the field contaminants and cluster members in the other regions of CMD. In Sec.~\ref{s:binaries}, 
we further restricted our sample of cluster members by selecting stars close
to the main-sequence fiducial line
and to the equal mass binary sequence.

\subsection{Blue Stragglers}

\begin{figure}
\center
\includegraphics[width=8cm]{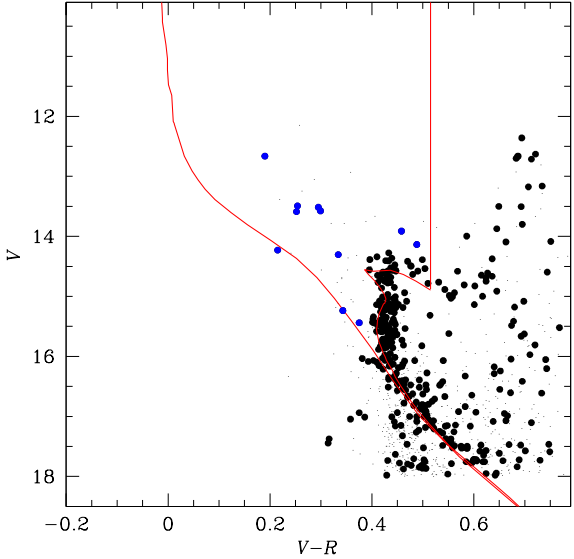}
    \caption{A $(V,V-R)$ CMD showing the area occupied by the
       blue straggler candidates. The blue
       dots indicate      the   selected  candidates of  blue stragglers
      listed      in        Table~\ref{tab:bluestr}.
       The big       black        dots   show      the       stars    that meet
       our         cluster member       selection        criteria,                whereas
      the small    black     dots  are the Galactic disk stars as indicated
by their proper motions.
        }
\label{fig:bluestr}
\end{figure}

Although \object{NGC~6253} is projected against a rich Galactic disk
stellar field, Fig.~7 clearly shows a conspicuous extension of the
cluster's main sequence brighter than the turnoff point, apparently
not related to Galactic disk.
 As emphasized by Carraro et al. (2008),
generally the Galactic disk main sequences occupy the same region where
the blue straggler stars are routinely searched for.
If the kinematic membership information is not known, this bears the
consequences on the true number of blue straggler among the candidates
and on the interpretation of their statistics and formation scenario (De Marchi et al. 2006). \\
Ahumada \& Lapasset (2007) suggested that \object{NGC~6253} harbors
27 blue stragglers and it is  one of the  richer star clusters in terms of
blue stragglers. To estimate the number of cluster's blue stragglers,
we  followed   the same method as applied by Ahumada \& Lapasset (2007).
Thus, in      Fig.~\ref{fig:bluestr}  the  best fitting Padova isochrone
for the $(V,V-R)$ CMD is plotted along with the Zero Age Main Sequence,
adjusted to the same metallicity as the cluster. The triangular area
between the ZAMS and a certain portion of theoretical isochrone
defines  the area containing blue stragglers.
As to the red limit of this area, we applied a
cut-off at $V-R\sim0.5$, denoted by the vertical line in Fig.~\ref{fig:bluestr}.
The stars that passed our cluster member selection criteria
are shown as large black dots. The stars considered to be the blue stragglers
candidates are indicated by large blue dots. The likely Galactic disk
contaminants in the same region and magnitude range are
plotted as small dots. Summarizing, the total number of
blue straggler candidates is 11 (Tab.~\ref{tab:bluestr}).  All these stars
have proper-motion membership       probability       P$>$91\%.
In this sample we  also           considered two stars
which are slightly on the blue side of our selection region, in accordance
with the  Ahumada \& Lapasset (2007) selection criteria.
In the same region of the CMD we can find another 27  stars that do not
meat our stringent selection criteria. This confirms the Carraro et al. (2008)
main conclusion that the statistics of  Ahumada \& Lapasset (2007)
is significantly  biased due to the field star contamination.

\subsection{Binary fraction}
\label{s:binaries}
A preliminary   estimate          of cluster's     binary    fraction
can         be    obtained   using        the radial velocity distribution
from previous spectroscopic studies combined with our proper-motion
membership probabilities. Thus, from  the sample of 11 probable
cluster members analyzed by Sestito et al.~(2007) and  Carretta et al.~(2007),
3 stars are probable binary stars, as indicated by their deviant
radial velocities and  high membership probability
(Sect.~\ref{s:spectroscopy}). This yields a binary fraction of $\sim\,30\%$.
Given the small number of cluster members, this
independent estimate has a high uncertainty and is applicable only
to the population of evolved stars in \object{NGC~6253}.

\begin{figure}
\center
\includegraphics[width=8cm]{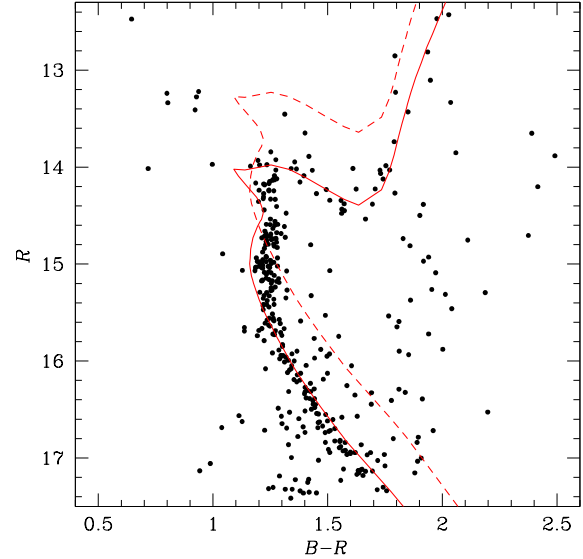}
\includegraphics[width=8cm]{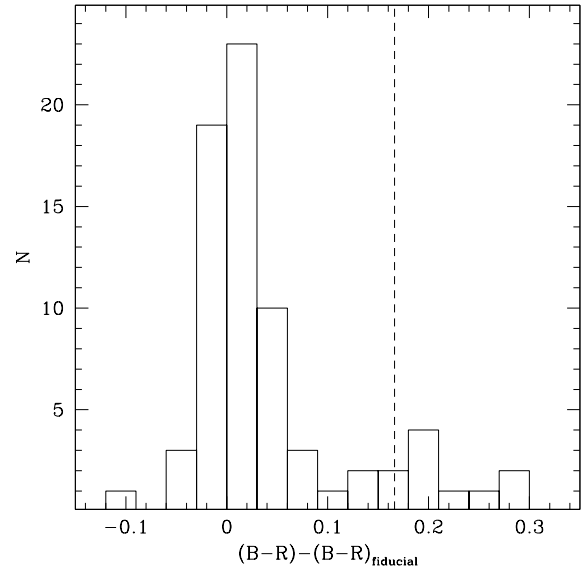}
    \caption{ {\it Upper panel:} $(R,B-R)$ CMD with a superimposed best-fit
     Padova isochrone (solid line), and the
     same isochrone shifted by 0.75 mag (dashed line), indicating
     the expected position of the equal mass binaries sequence. 
     {\it Bottom panel:} 
     histogram of color differences for stars in the magnitude range
     $15.3<R<16.3$ and color range $1.2<B-R<1.7$ with respect to the
     color of the main-sequence fiducial line (see text). The dashed
     line indicates the mean color difference between the stars located
     on the equal mass binary sequence and on the main sequence.
        }
\label{fig:rbr}
\end{figure}

Our proper-motion-selected sample of cluster members can be used 
to give an estimate of the cluster's binary fraction. In
Fig.~\ref{fig:rbr} (upper panel), we show the $(R,B-R)$ CMD with the
best-fit Padova isochrone overplotted (solid line) and the same
isochrone shifted by 0.75 mag (dashed line), which denotes the locus
of equal mass binaries. As evident from this figure, in the magnitude
range $15<R<16.5$ some stars appear to fall close 
to the predicted binary sequence, or in between the main sequence and
the binary sequence. The stars with color $B-R>1.7$ in the same
magnitude range are clearly field contaminants with proper motions similar to
those of cluster stars. At $R<15$
 the main sequence and the equal-mass binary sequence
overlap, but at $R>16.5$ the binary sequence and the field contaminants
are not anymore discernible. Moreover, for $R>16.5$ some clear contaminants 
are also present on the blue side of the main sequence. Finally, the stars
with similar proper motions to cluster members which are just located on
the main sequence or on the binary sequence cannot be isolated with the
present data. Similarly, the color-color diagram turned out
to be inappropriate for this purpose (Sec.~\ref{s:twocolors}), although
one can argue that these field contaminants affect equally both the single
cluster members and cluster binary stars and thus, have
a minor impact on the fraction of binaries relative to the total
number of cluster members. It should be noted that the binaries stars with
mass ratio $q<0.5$ should appear on the main sequence
as their color and magnitude differences are smaller than the
photometric errors (e.g., Fan et al.~1996, BR). Then, the binary fraction
calculated below can be considered as a crude lower limit of its true value.

Using the diagram in Fig.~\ref{fig:rbr} (upper panel) we isolated stars in the magnitude range
$15.3<R<16.3$ and with colors $1.2<B-R<1.7$. We then considered the manually-selected main-sequence fiducial line in that magnitude range and derived
the histogram
of the color differences of all stars with respect to the colors of fiducial line
(Fig.~\ref{fig:rbr}, bottom panel). A secondary peak located
approximately at the  mean color predicted for equal mass binaries
 ($B-R\sim0.17$, vertical dashed line) 
is visible. In this histogram the number of stars with color $B-R>0.1$
is $13$ whereas the total number stars is $72$, indicating that the fraction
of binary stars should be $>18\%$ -- in good agreement with BR which
found a similar binary fraction ($>20\%$) but without the benefit of
proper motions.

\begin{table}
\caption{
Blue straggler candidates in \object{NGC~6253}.
\label{tab:bluestr}
}

\begin{center}
\begin{tabular}{c c c c c c}
\hline
 ID$_{\rm MO}$ &     RA   &   DEC   & $V$ & $B-V$  &   P (\%)\\
           & (J2000) & (J2000) &     &        &        \\
\hline
  45497 & 254.784714  & -52.709400  &    14.230   &    0.502  & 94  \\
  45387 & 254.777725  & -52.723915  &    13.588   &    0.551  & 97  \\
  45367 & 254.826706  & -52.718845  &    13.493   &    0.545  & 96  \\
  45444 & 254.753860  & -52.714283  &    13.516   &    0.642  & 95  \\
  45447 & 254.786285  & -52.702099  &    13.575   &    0.629  & 97  \\
  45377 & 254.776123  & -52.753593  &    14.305   &    0.663  & 91  \\
  45422 & 254.771103  & -52.710304  &    12.663   &    0.455  & 93  \\
  45455 & 254.839020  & -52.712032  &    13.912   &    0.855  & 96  \\
  45465 & 254.756470  & -52.723072  &    14.136   &    0.913  & 97  \\
  40942 & 254.772125  & -52.645889  &    15.237   &    0.699  & 94  \\
  33095 & 254.745377  & -52.644726  &    15.440   &    0.753  & 93  \\
\hline
\end{tabular}
\end{center}
\end{table}

\section{The Catalog}
\label{s:catalog}
The catalog  containing   all   $187963$   stars is only available 
in electronic form at the CDS via anonymous ftp to 
cdsarc.u-strasbg.fr (130.79.128.5) or via http://cdsweb.u-strasbg.fr/cgi-bin/qcat?J/A+A/ .
In Table~\ref{tab:catalog}, we list only the content of the catalog. If a
particular parameter is not measured or determined, we use a default
meaningless value of $-$10000.

\section{Conclusions}
\label{s:conclusions}
We present    a       photometric         and        astrometric  catalog
of                $187963$                    stars  in  the   $BVRIJHK$ bands in the
field     of     the   old and  metal-rich    Galactic   open  cluster \object{NGC 6253}.
The catalog was constructed using the images taken with  the
La Silla 2.2m telescope and the Siding Spring 3.9m
telescope   during   a     joint $10$ days observing campaign in June 2004.

We derived proper motions for stars out to $\sim\,3\arcmin$ from the cluster
center, and used these data to improve our knowledge of fundamental
properties of \object{NGC~6253}.
We compared our new $BVRI$ photometry with previous photometric data
and concluded that overall our photometry is more consistent superior
over the existing studies.

Relative proper motions derived in this study combined with the broadband
optical and near-infrared
photometry allowed us a more detailed analysis of the cluster's fundamental
properties. By the means of isochrone fitting we derived the true distance
modulus $\rm (V_0-M_v)=11.1-11.3$ and the corresponding reddening estimates
to be $\rm E(B - V)\,=\,0.15$, $\rm E(V - I)\,=\,0.24-0.26$,
$\rm E(V - J)\,=\,0.5$ and   $\rm E(V - H)\,=\,0.5-0.6$. The age of the cluster we obtained
is older than what previously suggested  being  $\sim3.5$ Gyr.  The color excess ratios obtained using  both
optical and near-infrared    photometry   are in agreement with   a  normal reddening law. Finally, we estimated the cluster's binary fraction to be $\sim\,20\%-30\%$    and
identified $11$ blue straggler candidates. More extensive spectroscopic survey
of \object{NGC~6253} is clearly warranted in order to improve the cluster
membership and identify binaries in this astrophysically important open
cluster. 

\acknowledgements{
{We thank L.~R.~Bedin for sharing the near-infrared data
and J.~Anderson for having made available his routines for
calculating proper motions. I.~Platais gratefully acknowledges support from the
National Science Foundation through grant AST 09-08114 to Johns
Hopkins University. We thank the anonymous referee for careful reading
and useful suggestions. This research was supported by the DFG cluster of 
excellence 'Origin and Structure of the Universe' 
(www.universe-cluster.de).
}

\begin{table*}[!]
\caption{Content of the \object{NGC~6253} catalog.
\label{tab:catalog}
}
\begin{center}
\begin{tabular}{c l}
\hline
 Col.  1 & Sequential identification number\\
 Col.  2 & x coordinate in the CCD reference system (column 28)\\
 Col.  3 & y coordinate in the CCD reference system (column 28)\\
 Col.  4 & Right Ascension (J2000)\\
 Col.  5 & Declination (J2000)\\
 Col.  6 & Calibrated B magnitude\\
 Col.  7 & Error in B magnitude\\
 Col.  8 & Calibrated V magnitude\\
 Col.  9 & Error in V magnitude\\
 Col. 10 & Calibrated R magnitude\\
 Col. 11 & Error in R magnitude\\
 Col. 12 & Calibrated I magnitude\\
 Col. 13 & Error in I magnitude\\
 Col. 14 & Calibrated J magnitude\\
 Col. 15 & Error in J magnitude\\
 Col. 16 & Calibrated H magnitude\\
 Col. 17 & Error in H magnitude\\
 Col. 18 & Calibrated K magnitude\\
 Col. 19 & Error in K magnitude\\
 Col. 20 & J magnitude from $2MASS$\\
 Col. 21 & H magnitude from $2MASS$\\
 Col. 22 & K magnitude from $2MASS$\\
 Col. 23 & Relative proper motion in Right Ascension (mas/year)\\
 Col. 24 & Error in RA proper motion\\
 Col. 25 & Relative proper motion in Declination (mas/year)\\
 Col. 26 & Error in Dec proper motion\\
 Col. 27 & Membership probability\\
 Col. 28 & CCD number in the system of MPG/ESO WFI\\
\hline
\end{tabular}
\end{center}
\end{table*}

{}

\end{document}